\documentclass[preprint,12pt]{elsarticle}

\usepackage{amssymb}
\usepackage{amsmath}
\usepackage{bm}
\usepackage{color}
\usepackage{ulem}
\usepackage{mathrsfs}
\usepackage{comment}

\usepackage{algorithm}
\usepackage{algpseudocode}

\newcommand*\kai[1]{\textcolor{red}{#1}}

\journal{Journal of Computational Physics}

\newcommand{\Keigo}[1]{\textcolor{blue}{#1}}
\newcommand{\keigo}[1]{\textcolor{blue}{#1}}

\begin{document}

\begin{frontmatter}



\title{Multiresolution analysis on tessellation graphs for inertial particle dynamics}


\author[JAMSTEC]{Keigo Matsuda\corref{cor1}} 
\affiliation[JAMSTEC]{organization={
Research Institute for Earth and Information Sciences, 
Japan Agency for Marine-Earth Science and Technology (JAMSTEC)},
            addressline={3173-25 Showa-machi, Kanazawa-ku}, 
            city={Yokohama},
            postcode={236-0001}, 
            state={Kanagawa},
            country={Japan}}
\cortext[cor1]{k.matsuda@jamstec.go.jp}

\author[AMU,JAMSTEC]{Thibault Maurel--Oujia} 
\author[AMU]{Kai Schneider} 
\affiliation[AMU]{organization={Institut de Mathématiques de Marseille, Aix-Marseille Université, CNRS},
            addressline={3 place Victor Hugo, Case 19}, 
            city={Marseille},
            postcode={13331},
            country={France}}

\begin{abstract}
A multiresolution technique on tessellation graphs for particle dynamics is proposed. 
This allows to split spatial field data given on millions of discrete particle positions 
into scale-dependent contributions. 
The Delaunay tessellation is used to define the graph, and Voronoi cell volumes are used to satisfy volume conservation.
Our approach enables computation of the scale-dependent statistics of particle dynamics 
by leveraging a wavelet transformation of Lagrangian point particle data and is useful for characterizing particle clustering in turbulent flows. The technique is systematically verified by using synthetic data of randomly distributed particles in a two-dimensional plane. Then 
the applicability of the technique is demonstrated by extracting the scale-dependent particle velocity divergence of inertial particles in homogeneous isotropic turbulence from direct numerical simulation data. The result is verified by comparing the energy spectrum of the divergence with that obtained by a Fourier-based approach. 
Finally, the wavelet-based filtering to the particle velocity divergence is demonstrated to extract the effect of caustics in inertial particle clustering. 
\end{abstract}

\begin{graphicalabstract}
\includegraphics{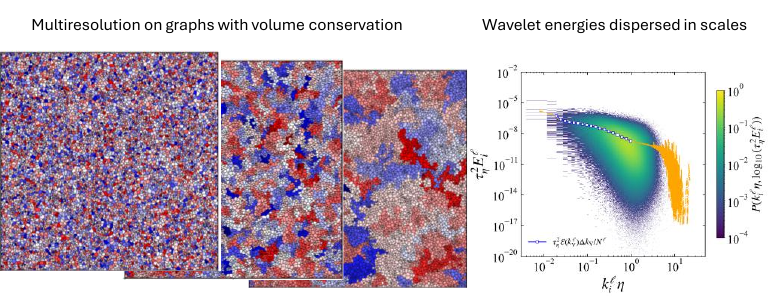} 
\end{graphicalabstract}

\begin{highlights}
\item 
A multiresolution technique on graphs 
to analyze dynamics of millions of particles.
\item 
Tessellations are used to generate graphs to represent physical fields.
\item 
The technique allows decomposition of unstructured data in scale and space.
\item 
Application
to scale-dependent statistics of particle clustering dynamics in turbulence. 
\item 
Filtering for the effect of caustics in inertial particle clustering is demonstrated.
\end{highlights}

\begin{keyword}
Multiresolution \sep Graph \sep Tessellation \sep Particle-laden flow \sep Clustering \sep Scale-dependent statistics 


\end{keyword}

\end{frontmatter}


\newpage

\section{Introduction}
\label{introduction}

Inertial heavy particles suspended in high Reynolds number turbulence are ubiquitous in geophysical and industrial flows; e.g., cloud droplets in atmospheric flows, dust particles in protoplanetary disks, and spray combustion. 
Inertial particles form nonuniform spatial distributions, which is referred to as inertial particle clustering, 
in turbulence because the inertial particle trajectory deviates from the fluid particle trajectory. 
When the particle response time is sufficiently smaller than the fluid flow time scale, the particle clustering is dominated by the preferential concentration mechanism \cite{Maxey1987,Squires}, in which the particles are swept out from turbulent eddies and concentrate in low vorticity and high strain rate regions, i.e., the particle concentration is determined by local flow structure. When the particle response time is larger than the flow time scale, the nonlocal clustering mechanism becomes dominant, and the particle path history affects the clustering formation \cite{Gustavsson&Mehlig2011,Bragg&Collins2014,Bragg2015}.
Recent reviews on particle-laden flow can be found in~\cite{brandt2022particle, bec2024statistical, marchioli2025particle}.

The particle clustering in high Reynolds number turbulence has multiscale structures. 
Recently, Matsuda et al.~\cite{matsuda2021scale} showed that the cluster or void pronounced structures of inertial particle clustering depend on the spatial scale and the Stokes number. 
The multiscale structures of clustering are observed likewise 
in the inertial subrange of high Reynolds number turbulence 
\cite{matsuda2024heavy}.
Therefore, to predict particle behavior, it is important to understand and model the multiscale dynamics of the particle clustering.
The key quantity to evaluate the particle cluster formation is the divergence of the particle velocity.
However, it is still challenging to obtain multiscale particle velocity divergence information based on discrete Lagrangian particle position and velocity data, i.e., particle cloud data.



Tools from discrete mathematics enable efficient description and representation of particle cloud data, 
for instance tessellations and graphs, 
which have been successfully used in various applications, e.g. in neuroscience and telecommunication, but also
for particle laden flows, e.g., for quantifying particle clustering, see \cite{monchaux2010preferential}.
Voronoi and Delaunay tessellations provide a geometrical  framework for defining neighborhood relationships on arbitrary unstructured grids. Voronoi tessellation has been widely applied to analyze data across diverse fields, including astrophysics, biology, and particle-laden turbulence, see e.g., \cite{ebeling1993detecting, monchaux2010preferential, obligado2014preferential}. Its first experimental application to turbulent inertial particle-laden flows was reported by Monchaux et al. \cite{monchaux2010preferential}. Since then, Voronoi-based methods have become increasingly popular for studying particle clustering in simulation data. For comprehensive reviews, see Monchaux et al.~\cite{monchaux2012analyzing} and Brandt and Coletti \cite{brandt2022particle}. The volume of a Voronoi cell directly reflects the local point density, enabling density statistics that are independent of arbitrary choices, such as bin size, commonly used in traditional density estimation methods.
Oujia et al.~\cite{oujiaJFM2020} proposed a tessellation technique to calculate the particle velocity divergence using the position and velocity of Lagrangian particles. This technique allows us to access the particle velocity divergence at particle position based on the temporal rate of change in volume of a tessellation cell.

The use of a graph representation of the particle data, i.e., a mathematical structure of vertices and edges,
permits then the application of signal processing techniques on graphs, a growing subject over the last years, see e.g. \cite{shuman2013emerging}.
In particular, multiresolution analysis, which is related to wavelets, can be constructed and applied to tessellations. Thus, tessellations can be coarse grained and refined to obtain scale information. 
Multiresolution constructions were originally developed on Cartesian grids \citep{mallat1999wavelet}, and generalizations for triangles can be found in \cite{cohen2000multiresolution} and \cite{yu1999fast}.
%
%
More recently, wavelets beyond Euclidean spaces have been developed, including wavelets on graphs.
%
%
For instance wavelets on discrete data clouds, graphs and manifolds 
where diffusion is used as a smoothing and scaling tool to enable coarse graining and multiscale analysis can be found in \cite{coifman2006diffusion}.
A spectral graph wavelet transform (SGWT) which can be seen as an extension of the continuous wavelet transform, was introduced in
\cite{hammond2011wavelets}.
Hereby the underlying idea is the spectral decomposition of the discrete graph Laplacian.
Some review on graph signal processing and tools for processing data defined on irregular graph domains, including some history can be found in
\cite{ortega2018graph}.
A graph multiresolution analysis for piecewise smooth graph signals is proposed in
\cite{chen2018multiresolution} using a coarse-to-fine approach, which iteratively partitions a graph into two subgraphs.
Multiresolution on finite connected weighted graphs using random forests has been proposed in~\cite{avena2020intertwining}.
%


In the current paper, 
we develop a straightforward approach for a multiresolution technique on graphs to analyze field data bundled on particle clouds, having the multiscale decomposition of the particle velocity divergence. 
To this end, a Delaunay tessellation is utilized for the graph, and a Voronoi tessellation is used to define the relationship between the graph and the field.
A first version of this approach together with preliminary results can be found in \citep{matsuda2022multiresolution}.
Note, that in our approach we perform the multiresolution analysis going from fine to coarse scale, while classically coarse to fine strategies are adopted.
Coarsening and refinement operators are introduced, and wavelet coefficients are defined as details between two levels. 
A fast algorithm for this purpose is proposed, which works efficiently for millions of particles, and thus scale decomposition of the particle velocity divergence can be determined, yielding insight into the multiscale dynamics of clustering. 
%

The remainder of this paper is organized as follows. 
In Section~\ref{particle-laden_flow}, 
the basic equations for the considered particle-laden flows and the tessellation technique to analyze the particle velocity divergence are briefly outlined. 
The proposed multiresolution technique is described in Section~\ref{multiresolution}. 
Sections~\ref{app2d} and \ref{app3d} present the results from application of the proposed technique to two-dimensional (2D) data of random particles with artificial test signals and three-dimensional (3D) data of inertial particles with the particle velocity divergence data in homogeneous isotropic turbulence (HIT), respectively.
Finally, the conclusions are drawn in Section~\ref{conclusion} including some perspectives for future applications. 
The appendices present technical details on the link between discrete multiresolution and biorthogonal wavelets, and normalization issues.
\section{Particle-laden flows and tesselations}
\label{particle-laden_flow}

\subsection{Governing equations}

    We consider the Navier--Stokes equations
    for an incompressible flow,
    \begin{equation}
        \frac{\partial {\bm u}}{\partial t} + ({\bm u}\cdot\nabla){\bm u} = -\frac{1}{\rho}\nabla p + \nu \nabla^2 {\bm u} + {\bm f} \; ,
\quad \quad
        \nabla\cdot{\bm u} = 0
        \label{eq:navierstokes},
    \end{equation}
%
where ${\bm u}$ is the fluid velocity field, $p$ is the pressure, $\rho$ is the fluid density, $\nu$ is the kinematic viscosity, and ${\bm f}$ is an external forcing.
We consider an $m$-dimensional cubic domain $\Omega \subset \mathbb{R}^m$ (with $m=2$ or $3$ here) of side length $2 \pi$ and periodic boundary conditions, i.e., a torus $\mathbb{T}^m$ where $\mathbb{T} = 2 \pi (\mathbb{R} / {\mathbb{Z}})$; the equations are completed with suitable initial conditions.
%
%
The Lagrangian motion of small and heavy inertial particles in a fluid flow can be modeled by the following equations based on the point particle assumption,
\begin{equation}
    \frac{d {\bm x}_p}{d t} = {\bm v}_p \; , \quad \quad 
    \frac{d {\bm v}_p}{d t} = -\frac{{\bm v_p}-{\bm u}({\bm x}_p)}{\tau_p}
    \label{eq:particle},
\end{equation}
where ${\bm x}_p$ is the particle position, ${\bm v}_p$ is the particle velocity, and $\tau_p$ is the particle relaxation time. The term on the right-hand side of the second equation in Eq. \eqref{eq:particle} is the drag force term.
Here, we assume that the particles density is sufficiently larger than the fluid density, and that the effect of gravitational settling is negligibly small. 
When the particles are moving in a turbulent flow, statistics of the particle clustering depend on the Stokes number $St$ defined as the ratio of $\tau_p$ and a flow time scale. 
The particles are initially uniformly distributed, initialized with the fluid velocity  and injected into a statistically stationary flow. 
Note that the particles are one-way coupled and thus do not retroact on the carrier flow.  Details on the numerics and physical model can be found in  \cite{Matsuda2014} and \cite{Maxey1987}.

\subsection{Particle velocity divergence}
    The particle clustering mechanism can be discussed using the conservation equation of the particle number density field $n$ in a continuous setting
    considering the ensemble average of possible realizations of particle distributions in the same turbulent flow.
    If there is no external source or sink for the particle number density, the conservation equation is given by
    \begin{equation}
        \frac{\partial n}{\partial t} + {\bm v}\cdot\nabla n = -n \nabla\cdot{\bm v},
    \end{equation}
    where ${\bm v}$ is the particle velocity field in a continuous setting. 
    This means that the source and sink for the number density $n$ along the Lagrangian particle trajectory are given by the term on the right-hand side, and the negative/positive values of the particle velocity divergence $\nabla\cdot{\bm v}$ contribute to the cluster formation/destruction, respectively.
    If the particles are passive tracers, which perfectly follow the fluid motion in an incompressible flow, $\nabla\cdot{\bm u} = 0$, which is volume preserving,
    the particle velocity divergence $\nabla\cdot{\bm v}$ becomes likewise  zero and particle clustering does not occur.
    Thus, the source of clustering is the deviation of particle motion from the fluid flow.
    Maxey \cite{Maxey1987} evaluated the particle velocity divergence using an approximation for 
    $St \ll 1$. 
    Maxey's approximation yields that $\nabla\cdot{\bm v}$ is proportional to the second invariant of the velocity gradient tensor, and it explains the preferential concentration mechanism. 
    However, the approximation is not valid for particles with 
    $St \gtrsim 1$, which show significant particle clustering. 

\subsection{Tessellation technique}
\label{sec:tess_tech}

    The particle distribution and velocity data can be obtained by performing a direct numerical simulation (DNS) of particle-laden flows, in which the fluid flow field ${\bm u}$ is obtained by solving Eqs.~(\ref{eq:navierstokes}) 
    on Eulerian grid points and the particle position and velocity are obtained by solving Eqs.~(\ref{eq:particle}) based on the Lagrangian method.
    The difficulty in calculating the particle velocity divergence is due to the discrete nature of the particles and the corresponding velocity. 

The Voronoi tessellation (or diagram) is a technique to construct a decomposition of the space, i.e., the fluid domain $\Omega$, into a finite number of Voronoi cells $\{\Omega_0,\dots, \Omega_{N_p-1}\}$. 
When a finite number of particles 
$\{P_0,\dots, P_{N_p-1}\}$
are dispersed in space, a Voronoi cell $\Omega_i$ is defined as a region closer to a particle than other particles.
Note that the union of all cells is equal to the fluid domain, i.e. 
$\bigcup_{i=0}^{N_p-1} \Omega_i = \Omega$, and that any non-empty intersection between two cells defines the common boundary of these two cells.
The volume of a Voronoi cell is referred to as Voronoi volume and denoted by $V_i$. 
The  cell $\Omega_i$ can be interpreted as the zone of influence of the particle $P_i$. 
The larger the number of particles in a given domain,
the smaller the Voronoi volume.
The diagram will allow us to identify particles inside clusters (corresponding to small cells) and particles inside void regions (corresponding to large cells).
A survey on Voronoi diagrams, a classical technique in computational geometry, can be found in 
\cite{aurenhammer1991voronoi}.

To compute the particle velocity divergence based on the DNS data, 
we proposed a tessellation technique based on the temporal rate of change in volumes \cite{oujiaJFM2020}.  
The discrete particle velocity divergence 
${\cal D}({\bm x}_{p,i})\approx(\nabla\cdot{\bm v})({\bm x}_{p,i})$
at the particle position 
${\bm x}_{p,i}$ is given by
\begin{equation}
    {\cal D}({\bm x}_{p,i}) = \frac{1}{V_i}\frac{D V_i}{D t}
    \label{eq:Div},
\end{equation}
where 
$V_i$ is the volume of the tessellation cell that contains the particle at position ${\bm x}_{p,i}$. 
To define the tessellation cells around a particle, 
the 
Voronoi tessellation is used in \cite{oujiaJFM2020}, and recently in \cite{MaurelOujia2024JCP} we used the center of gravity of Delaunay tetrahedron for the vertex of the cell for stability reasons. 
Periodic boundary conditions are taken into account in the tessellation procedure when the computational domain is periodic. To this end, particles located near one boundary are replicated to the halo region on the opposite boundary, producing ghost particles, as if the domain were periodically repeated in space, and the tessellation is constructed on a particle set extended for the halo regions.
The tessellation technique allows us to access the particle divergence values at the particle positions. 
The technique has been applied to particle clustering in homogeneous isotropic turbulence \citep{oujiaJFM2020,MaurelOujia2024JCP}, turbulent flows in porous media \cite{Apte2022JFM} and channel flows \citep{West2022CTR,West2024IJMF}.
However, to analyze the multiscale clustering dynamics, it is necessary to develop a multiresolution tessellation technique for the particle velocity divergence data.

\section{Multiresolution analysis on graphs}
\label{multiresolution}


\subsection{A primer on discrete multiresolution on regular grids}
\begin{figure}
    \centering
    \includegraphics[width=0.9\linewidth]{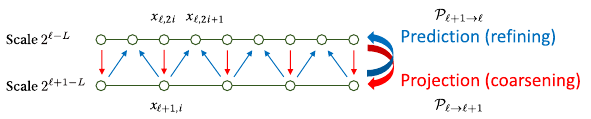}
    \caption{Illustration of projection and prediction operators for 1D regular grids. 
    }
    \label{fig:1Dcase}
\end{figure}

The discrete multiresolution analysis for point values and cell averages has been introduced by Harten in the context of adaptive finite difference and volume schemes \cite{harten1993discrete,harten1995multiresolution, harten1996multiresolution}.
Here we consider a discrete signal 
$\{s_n\}=\{\bar{s}^{\ell=0}_{n}\} \in \mathbb{R}^N$ on one-dimensional 
(1D) equidistant grids at $\{x^{\ell=0}_{n}\} \in \mathbb{R}^N$, where $N=2^L$ and the scale of the grid spacing is $2^{-L}$. 
The level $\ell = 0$ is assigned to the finest grid while $\ell = L$ corresponds to the coarsest.
Note that the signal represents a set of scalar-valued data, either cell averages or point values on discrete points.
We recall that the signal can be a spatial field data sampled at discrete points distributed in the space.
For the multiresolution analysis, the signal at level $\ell$ given on $2^{L-\ell}$ grid points is projected onto level $\ell+1$ with $2^{L-\ell-1}$ grid points by coarsening the grid, $x^{\ell+1}_{i}=x^\ell_{2i}$ for $i=0,\cdots,2^{L-\ell-1}-1$. 
Hence, the scale of grid spacing becomes $2^{-(L-\ell-1)}$.
The coarsening of the signal at level $\ell$ can thus be described by
\begin{equation}
    \bar{s}^{\ell+1}_{i} = {\cal P}_{\ell \to \ell+1}\{\bar{s}^\ell_{n}\}
    \label{eq:projection}, 
\end{equation}
where $\bar{s}^{\ell+1}_{i}$ is the signal at $x^{\ell+1}_{i}$ at the coarser scale, and ${\cal P}_{\ell \to \ell+1}$ is the projection operator. 
By applying the projection operator to the coarser signal recursively, we can define the signal $\{\bar{s}^\ell_{n}\}$ at the scale $2^{\ell-L}$ ($\ell=0,\cdots,L$), where the level $\ell$ increases as the scale becomes coarser.
Figure \ref{fig:1Dcase} shows the relationship between the scale $2^{\ell-L}$ and $2^{\ell+1-L}$.
The signal $\{\bar{s}^\ell_{n}\}$ at the scale $2^{\ell-L}$ can thus be projected onto the scale $2^{\ell+1-L}$ based on Eq.~(\ref{eq:projection}). 
The signal at the finer scale $2^{\ell-L}$ can be predicted by using the signal at the coarser scale $2^{\ell+1-L}$, 
\begin{equation}
    \hat{s}^\ell_{2i+1} = {\cal P}_{\ell+1 \to l}\{\bar{s}^{\ell+1}_{n}\}
    \label{eq:prediction}, 
\end{equation}
where $\hat{s}^\ell_{i}$ is the predicted signal at $x^\ell_{i}$, and ${\cal P}_{\ell+1 \to l}$ is the prediction operator.
The detail information confined in space and scale can then be defined as the difference between the original and the predicted signal,
\begin{equation}
    d^{\ell+1}_{i} = \bar{s}^\ell_{2i+1} - \hat{s}^\ell_{2i+1},
\end{equation}
where $d^\ell_{i}$ is the detail coefficient for the level $\ell$ and the position index $i$. 
The projection and prediction operators determine the characteristics of the multiresolution analysis. 
For the wavelet transform, the operators must satisfy, 
\begin{equation}
    {\cal P}_{\ell \to \ell+1} \circ {\cal P}_{\ell+1 \to l} = Id
    \label{eq:identity},
\end{equation}
where $\circ$ is the composition operator and $Id$ the identity.
This means that applying the projection to the prediction does not generate any information.
%
\footnote{For regular grids with $N=2^J$ grid points, one typically uses the scale index $j$ which is related to the level $\ell$ via $j = L -l$. Note that the coarsest scale index $j=0$ corresponds to the level $L$ and the finest scale index $j=J$ to the level $\ell=0$. The scale with scale index $j$ or level $\ell$ is then given by $2^{-j} = 2^{\ell - L}$. Note that here we use $L=J$.}

\subsection{Multiresolution graphs}



To construct a multiresolution technique for a signal on discrete particle positions, it is necessary to define the projection and prediction operators, and these operators require the information of neighbor particles.
Discrete particle points and particle pairs that indicate neighbors 
constitute an undirected graph. 
In mathematics, a simple undirected graph $G$ is defined as a pair $G = (P,Q)$, where $P$ is a set of vertices and $Q$ is a set of edges defined as pairs $\{P_0, P_1\}$ of elements of $P$. 
Therefore, the projection and prediction operators are defined on a graph, in which particles are assigned as vertices.

We propose utilizing the Delaunay tessellation (diagram) to define the edges of the graph considering particles in the domain $\Omega \subset \mathbb{R}^m$ with $m = 2$ or 3. The Delaunay tessellation is dual of the Voronoi tessellation and uniquely determined by a set of particle positions. 
The Voronoi cell for each particle and the adjacent cells are connected by the edges of the Delaunay tessellation. 
Each vertex on the graph is linked to the physical space through the particle position ${\bm x}_{p,i}$ and the volume $V_i$ of the Voronoi cell.
The use of Delaunay tessellation is also compatible to the tessellation-based analysis for the particle velocity divergence.

Periodic boundary conditions are considered likewise for the multiresolution. The Delaunay tessellation 
is constructed in the same manner as written in Section \ref{sec:tess_tech}.
To construct a periodic graph, the resulting Delaunay connectivity is 
mapped back to the original domain by identifying each periodic image with its corresponding particle (vertex). Hence, connections established through 
the ghost particles (vertices) are replaced by the equivalent connections between particles (vertices) in the physical domain.

\begin{figure}
    \centering
    (a)\includegraphics[width=0.45\linewidth]{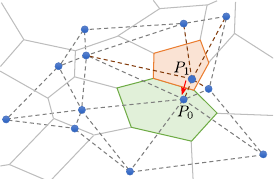}
    (b)\includegraphics[width=0.45\linewidth]{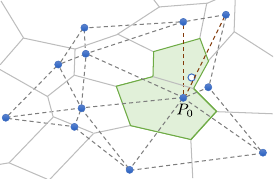}
    \caption{Point merging process on the Delaunay graph in 2D. (a) The graph before merging. The edges of the Delaunay graph represent 
    pairs of adjacent cells. The odd point $P_1$ will be merged to the even point $P_0$. 
    (b) The graph after merging. The graph edges are modified to represent connections with the cells 
    adjacent to the merged cell.
    }
    \label{fig:point_merge}
\end{figure}

To apply multiresolution analysis on graphs, multiresolution graphs are needed. 
Here, the multiresolution graph construction starts from the finest scale, which is level 0. Therefore, we use the level index $\ell=0,\cdots,L$, 
which increases as the scale becomes larger. 
Note that the constructed graphs do not change even when increasing the number of levels $L$. 
The coarser graph at level $\ell+1$ is constructed from the finer graph at level $\ell$. 
For the graph coarsening, we adopted the `half-edge collapse' operator \citep{kobbelt1998interactive}, in which one vertex is merged with an adjacent vertex.
Figure \ref{fig:point_merge} shows the schematic diagram of the merging process. 
Here, the vertex $P_1$ is merged to the vertex $P_0$, 
and $P_1$ is removed. The edges connected to 
$P_0$ and $P_1$ are integrated.
Based on the analogy to the 
1D regular grid case, 
$P_0$ and $P_1$ can be considered as the `even' and `odd' vertices. 
Since we consider particle distributions in the physical space, we assume that each particle (i.e., each vertex on the graph) has the information of the tessellation cell volume. 
We propose a coarser-graph construction procedure by finding a pair of vertices with the smallest volumes for the merging process, expecting that the vertex pair is relatively close each other and belongs to the same cluster of particles.
%
%
The algorithm for constructing the graph at level $\ell+1$ from the graph at level $\ell$ 
consists in successively merging, when possible, each cell, at most once, with one of its neighbors \cite{Dhillon2007}. 
The specific details are given in Algorithm~\ref{alg:graph_construction}.
%
Note that this is not a unique procedure to construct coarser graphs. There are several methods that use edge weights to determine the collapse edge \cite{Dhillon2007}. The advantage of the present procedure is that it does not require the edge weight, and thus it is memory saving.

\begin{algorithm}
\caption{Construction of the graph at level $\ell+1$ from level $\ell$}\label{alg:graph_construction}
\begin{algorithmic}[1]
\State Initialize all vertices in the graph at level $\ell$ as unmarked
\While{there is one or more unmarked vertices}
    \State Select the unmarked vertex with the minimum volume, and denote it as the odd vertex.
    \If{there is one or more unmarked vertices adjacent to the odd vertex}
        \State Select the vertex with the minimum volume among the unmarked adjacent vertices, and denote it as the even vertex.
        \State Merge the odd vertex to the even vertex: transfer all edges from the odd vertex to the even vertex, and remove the odd vertex. 
        \State Mark the even vertex to be shifted to level $\ell+1$.
    \Else
        \State Mark the odd vertex to be shifted to level $\ell+1$ without merging.
    \EndIf
\EndWhile
\end{algorithmic}
\end{algorithm}

The volume conservation is a key feature to represent the physics on the graph. 
Therefore, the volume value of the marked vertex is updated to the sum of the volumes attributed to the even and odd vertices.  
When the volume attributed to the even and odd vertices at level $\ell$ are given by $V_{2i}^{\ell}$ and $V_{2i+1}^{\ell}$, respectively, the volume attributed to the vertex marked and shifted to level $\ell+1$ satisfies 
$V_i^{\ell+1} = V_{2i}^{\ell} + V_{2i+1}^{\ell}$.
If the odd vertex is shifted to level $\ell+1$ without the merging process, the volume value is not updated.

    \subsection{Wavelet transform on graphs}




Here, we consider a scalar-valued signal $s_i^0=s^0({\bm x}_{p,i})$ ($i=0,\cdots,N_p-1$) on discrete particle positions (i.e., vertices on a graph).
Since we aim to apply the multiresolution analysis to field data such as the particle velocity divergence, 
we propose the wavelet transform on graphs based on the following conservation equation 
\begin{equation}
    V_i^{\ell+1} \bar{s}_i^{\ell+1} = V_{2i}^{\ell} \bar{s}_{2i}^{\ell} +V_{2i+1}^{\ell} \bar{s}_{2i+1}^{\ell}
    \label{eq:MRconservation},
\end{equation}
where $\bar{s}_i^\ell$ is the signal of the $i$th vertex on the graph at the level $\ell$, and the index $2i$ and $2i+1$ represent the even and odd vertices, respectively.
The projection operator is then defined as 
\begin{equation}
    \bar{s}_i^{\ell+1} = \frac{1}{V_i^{\ell+1} }(V_{2i}^{\ell} \bar{s}_{2i}^{\ell} +V_{2i+1}^{\ell} \bar{s}_{2i+1}^{\ell}) 
    \label{eq:MRprojection}.
\end{equation}
For the prediction operator, we simply assume that the predicted values are the same as the projected signals at the coarser level,
\begin{equation}
    \hat{s}_{2i}^{\ell} = \bar{s}_{i}^{\ell+1},  \, \, 
    \hat{s}_{2i+1}^{\ell} = \bar{s}_{i}^{\ell+1}.
\end{equation}
The prediction and projection operators satisfy the identity (\ref{eq:identity}).
The detail coefficients (i.e., the wavelet coefficients) are then given by the difference between the original and predicted signals, yielding 
\begin{equation}
    d_{i}^{\ell+1} = \frac{V_{2i}^{\ell}}{V_{i}^{\ell+1}}(\bar{s}_{2i+1}^{\ell} - \bar{s}_{2i}^{\ell}).
    \label{eq:MRdetails}
\end{equation}
The wavelet coefficient based on the $L^2$ normalization can be obtained  
by multiplying the detail coefficients with the scaling factor $\sigma_i^{\ell+1}=\sqrt{V_i^{\ell+1} V_{2i+1}^\ell/V_{2i}^\ell}$.
%
The relation of Harten's discrete multiresolution analysis \cite{harten1993discrete, harten1995multiresolution, harten1996multiresolution} to continuous biorthogonal wavelets~\cite{cohen1992biorthogonal} is detailed in \ref{app1}. 
A detailed derivation of the scaling factor is also explained in \ref{app2}.

\subsection{Filtering on graphs}
Using the detail coefficients, we can define filtering of the field on graphs.
The field $s^0({\bm x})$ is described as 
\begin{equation}
    s^0({\bm x}) = \overline{s}^L({\bm x}) + \sum_{\ell=1}^{L} \check{s}^\ell({\bm x})
    \label{eq:field_decomp},
\end{equation}
where $\overline{s}^L({\bm x})$ is the coarse-grained field at level $L$, and $\check{s}^\ell({\bm x})$ is the band-pass filtered field at level $\ell$ and given by
\begin{equation}
    \check{s}^\ell({\bm x}) = \sum_{i=0}^{N^\ell-1} d_i^\ell \psi_{i}^\ell({\bm x})
    \label{eq:field_bandpass}.
\end{equation}
Here $\psi_{i}^\ell({\bm x})$ is the wavelet with level index $\ell$ and position index $i$, and $N^\ell$ is the number of particles (vertices) at level $\ell$. Please see \ref{app1} for the details.
The low-pass filtered field, which is equivalent to the projected field, at level $\ell$ is then given by 
\begin{equation}
    \overline{s}^\ell({\bm x}) = \overline{s}^L({\bm x}) + \sum_{\ell'=\ell+1}^{L} \check{s}^{\ell'}({\bm x})
    \label{eq:field_lowpass}.
\end{equation}
Filtering on graphs can be defined more generally as 
\begin{equation}
    s^0({\bm x}) = \overline{s}^L({\bm x}) 
                 + \sum_{\ell=1}^{L} \sum_{i=0}^{N^\ell-1} f^\ell_i d_i^\ell \psi_{i}^\ell({\bm x})
    \label{eq:field_filtering},
\end{equation}
where $f^\ell_i \in [0,1]$ is an arbitrary filter. The low-pass and band-pass filtering are special cases where $f^\ell_i$ depends only on $\ell$. 

\subsection{Scale-dependent statistics on graphs}
Similarly to wavelets on Cartesian grids, see, e.g., \cite{farge2015wavelet}, we can define scale-dependent statistics on graphs using scale-dependent moments.
Spatial scales can be defined based on the volume information: the volume scale 
$d_V^\ell$ is defined as 
$d_V^\ell = [(1/N^\ell)\sum_{i=0}^{N^\ell-1} V_i^\ell/2]^{1/m}$,
where $V_{\rm mean} = (2\pi)^m/N_p$ is the mean volume, and $m$ is the space dimension. 
Note that the obtained $d_V^\ell$ can be approximated by $2^{\ell/m} V_{\rm mean}^{1/m}$.

For multiscale statistics of the signal on discrete particle positions, we can consider $q^{th}$ order level-dependent moments of details $d_{i}^{\ell}$,
\begin{equation}
    M_q[d_{i}^{\ell}] = \frac{1}{N^\ell}\sum_{i=0}^{N^\ell-1} (d_{i}^{\ell})^q,
    \label{eq:scaledependentmoments}
\end{equation}
where $N^\ell$ is the number of detail coefficients at level $\ell$.

Second order non-centered moments yield information of the energy at a given level $\ell$, also called scalogram in the wavelet literature \cite{farge1992wavelet}. Therewith, we can introduce a wavelet energy spectrum.
To this end,
we define the volume-based wavelength and wavenumber as $\lambda_V^\ell = 2 d_V^\ell$ and $k_V^\ell = \pi/d_V^\ell$, respectively, because the wavelength for each wavelet is given by the pair of adjacent volumes.
The wavelet energy spectrum is then defined as 
\begin{equation}
    {\cal E}(k_V^\ell) = \frac{1} {\Delta k_V^\ell} \sum_{i=0}^{N^\ell-1} E^\ell_i 
    = \frac{N^\ell M_2[\sigma_i^\ell d_i^\ell]} {(2\pi)^m \Delta k_V^\ell}
    \label{eq:MRenergyspectrum},
\end{equation}
where $E_i^\ell = (\sigma^\ell_i d^\ell_i)^2/(2\pi)^m$ is the wavelet energy for each detail coefficient, and
$\Delta k_V^\ell$ is the bandwidth of the wavelets at each level $\ell$ and approximately determined as 
$\Delta k_V^\ell = k_V^\ell \ln 2/m$.
The wavelet energy spectrum is defined based on the coefficient in the $L^2$ normalization by multiplying with the scaling factor $\sigma_i^\ell$ so that the spectrum has the same scaling as the Fourier spectrum.
Note that the wavenumber $k_V^\ell$ is the representative wavenumber of wavelets at level $\ell$, whereas it is not equivalent to the wavenumber in the Fourier analysis. As the volume $V^\ell_i$ is not unique even in the same level, the spatial scale of each wavelet is different from other wavelets.

Considering the spread of the scales of wavelets, we can define the volume-based wavenumber $k_i^\ell = \pi(2/V_i^\ell)^{1/m}$ for each wavelet. 
We can define a quasi-spherical energy spectrum by cumulating the wavelet energy over the same wavenumber bin of 
$k-1/2\le k_{{\rm V},i}^\ell < k+1/2$, i.e., 
\begin{equation}
{\cal E}'(k) = \sum_{k-1/2\le k_{{\rm V},i}^\ell < k+1/2} E^\ell_i.
\end{equation}
This wavenumber-based spectrum ${\cal E}'(k)$ represents the energy more confined in wavenumber compared with the level-based spectrum ${\cal E}(k_V^\ell)$, and thus this spectrum helps us understanding the relevance with the Fourier spectrum.

Another approach of taking scale-dependent statistics is level-based decomposition of the field in Eq.~(\ref{eq:field_decomp}).  
The moments of the band-pass filtered fields of Eq.~(\ref{eq:field_bandpass}) can be calculated by 
\begin{equation}
    {\cal M}_q[\check{s}^{\ell}] 
    = \frac{1}{V_\Omega}\sum_{i=0}^{N_p-1} \left\{ \check{s}^{\ell}({\bm x}_{p,i}) \right\}^q V_i
    \label{eq:bandpass_moments},
\end{equation}
where $V_\Omega = \sum_{i=0}^{N_p-1} V_i$ is the total volume of the domain
and $q$ the order of the moment.
The first- and second-order moments satisfy ${\cal M}_1[\check{s}^{\ell}]=0$ and ${\cal M}_2[\check{s}^{\ell}]=N^\ell M_2[\sigma^\ell_i d^\ell_i]/(2\pi)^m$, respectively.
Higher order scale-dependent statistics can be likewise defined using Eq.~(\ref{eq:bandpass_moments}). For example, scale-dependent flatness and skewness are given by ratios of moments, i.e., for flatness we have ${\cal F}^{\ell} = {\cal M}_4[\check{s}^{\ell}]/({\cal M}_2[\check{s}^{\ell}])^2$ and for skewness ${\cal S}^{\ell} = {\cal M}_3[\check{s}^{\ell}]/({\cal M}_2[\check{s}^{\ell}])^{3/2}$. These quantities allow to quantify the asymmetry of the probability distribution as different scales (e.g., if cluster or voids are dominating at different scales) and the departure from Gaussianity, important for quantifying intermittency of the clustering, without using projection onto Cartesian grids, as done in \cite{matsuda2021scale}.
\section{Applications}

\subsection{Two-dimensional test cases}
\label{app2d} 


The multiresolution tessellation is applied to random particle distributions in a 
2D square domain with edge length $2\pi$.  
The tessellation cells are defined by the standard 
Voronoi tessellation.
Two types of signals are generated for the signal value of each particle, instead of the particle velocity divergence data.
The first one is a Gaussian noise of ${\cal N}(0, 1)$, i.e., zero mean and unit variance. 
The second one is a spectral signal defined as
\begin{equation}
    s_i^0 = s^0({\bm x}_{p,i}) = A^{-1} \sum_{k_x=1}^{N_k}\sum_{k_y=1}^{N_k} a(|{\bm k}|) \sin[k_x x_{p,i} + \theta_x({\bm k})]\sin[k_y y_{p,i} + \theta_y({\bm k})], 
\end{equation}
where ${\bm x}_{p,i}=(x_{p,i}, y_{p,i})$, ${\bm k}=(k_x, k_y)$, $a(|{\bm k}|)=\exp[-|{\bm k}|^2 \pi^2/(24 k_c^2)]$ is the Gaussian filtered amplitude for the cutoff wavenumber of $k_c = 20$, and $\theta_x({\bm k}) \in [0,2\pi)$ and  $\theta_y({\bm k}) \in [0,2\pi)$ are uniformly distributed random phase shifts. The signal is normalized by $A$ so that the standard deviation of $s_i^0$ becomes unity. 
This signal is also Gaussian noise but spatially correlated due to the low-pass filtered amplitude. The Fourier energy spectrum of the signal is 
given by $2\pi k \{a(k)\}^2/(4A)^2$,
where $k=|{\bm k}|$. 
Thus, the spectrum is approximately proportional to $k$ for $k \ll k_c$ and has a peak at $k \approx 0.78 k_c$. 

\begin{figure}
    \centering
    \hspace{-2mm}
    \begin{minipage}{0.34\linewidth}
    (a)\\
    \includegraphics[width=\linewidth]{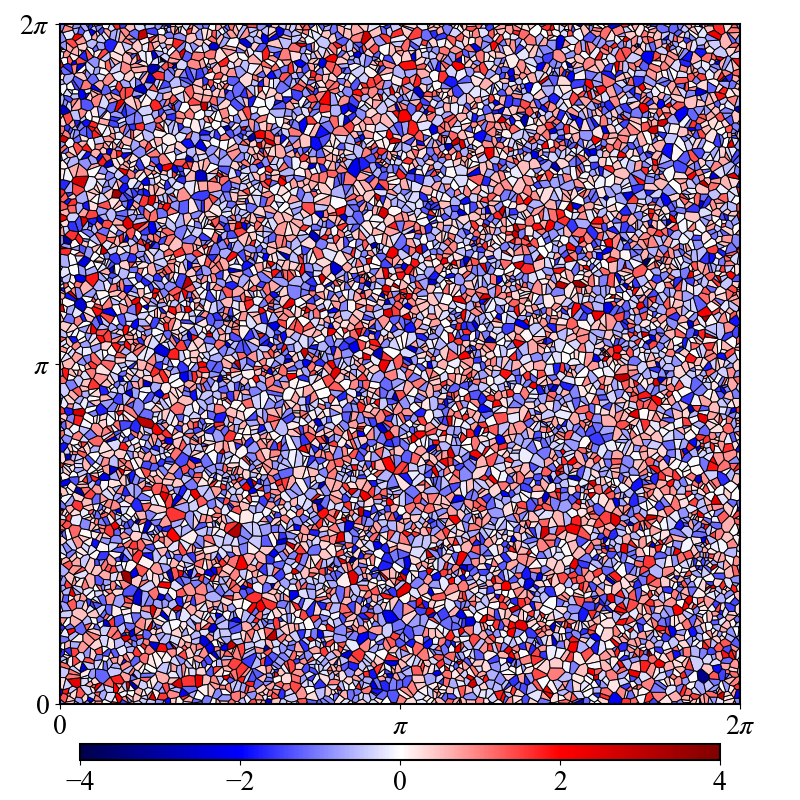}
    \end{minipage}
    \hspace{-5mm}
    \begin{minipage}{0.34\linewidth}
    (b)\\
    \includegraphics[width=\linewidth]{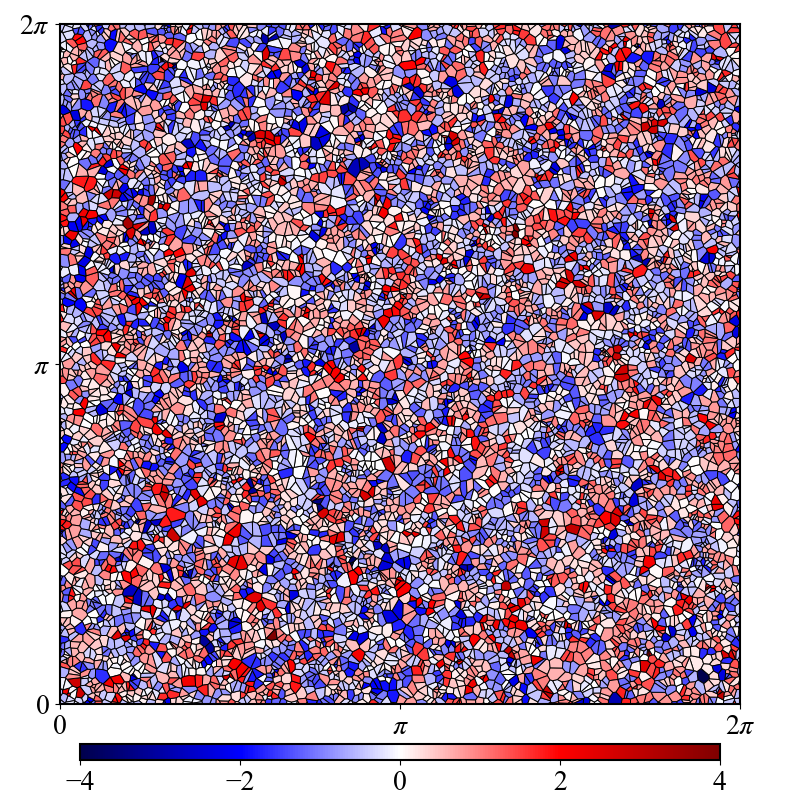}
    \end{minipage}
    \hspace{-5mm}
    \begin{minipage}{0.34\linewidth}
    (c)\\
    \includegraphics[width=\linewidth]{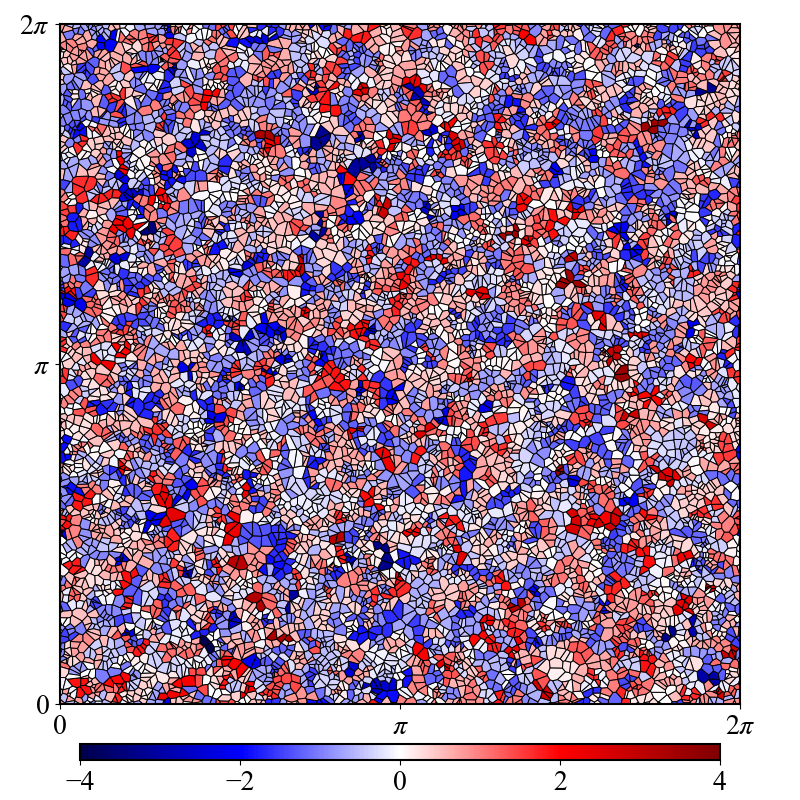}
    \end{minipage} \\
    \hspace{-2mm}
    \begin{minipage}{0.34\linewidth}
    (d)\\
    \includegraphics[width=\linewidth]{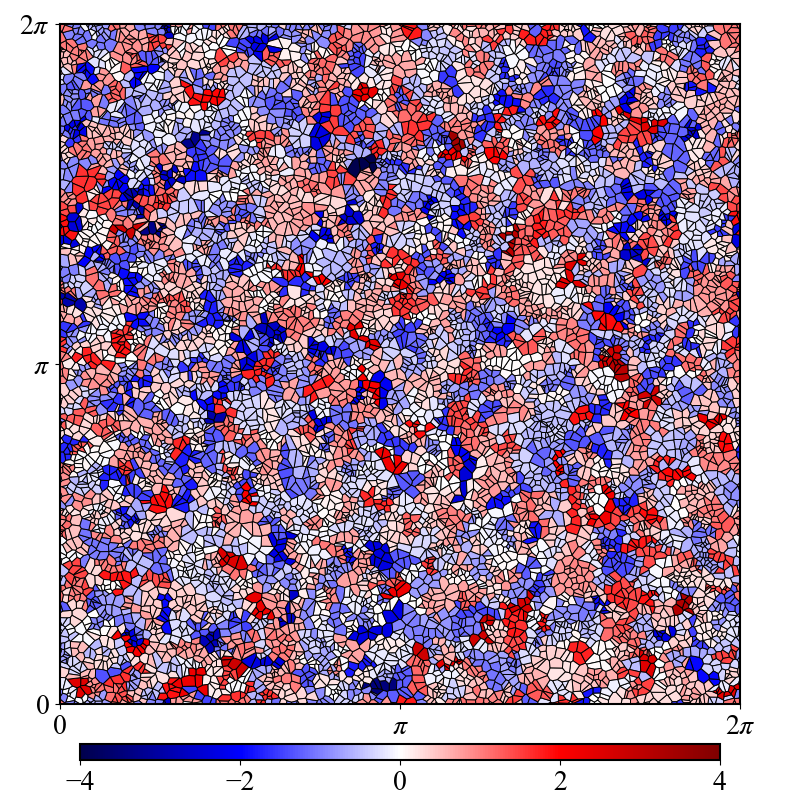}
    \end{minipage}
    \hspace{-5mm}
    \begin{minipage}{0.34\linewidth}
    (e)\\
    \includegraphics[width=\linewidth]{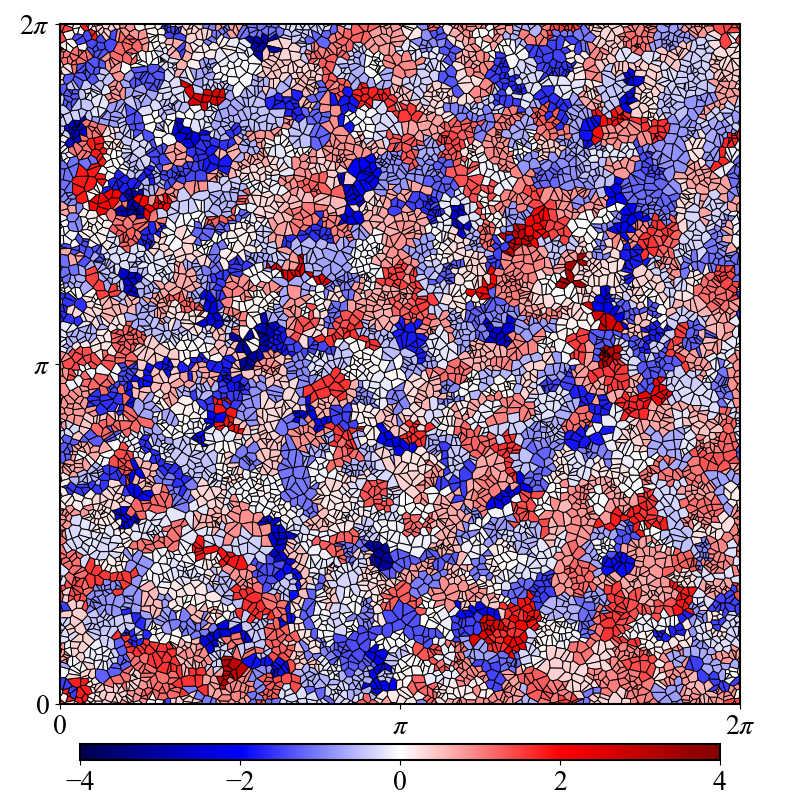}
    \end{minipage}
    \hspace{-5mm}
    \begin{minipage}{0.34\linewidth}
    (f)\\
    \includegraphics[width=\linewidth]{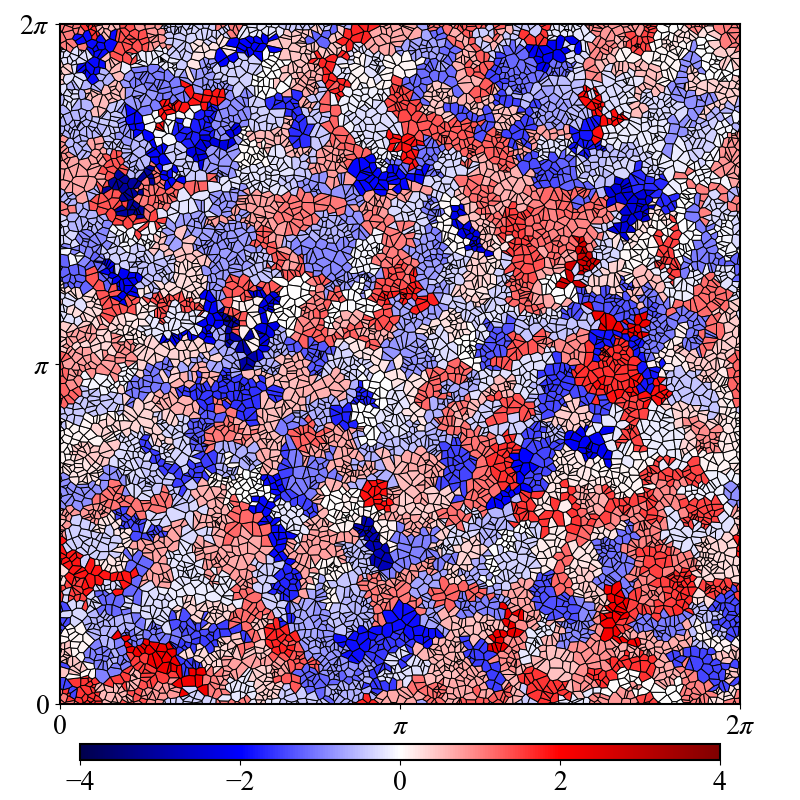}
    \end{minipage}
    \hspace{-2mm}
    \begin{minipage}{0.34\linewidth}
    (g)\\
    \includegraphics[width=\linewidth]{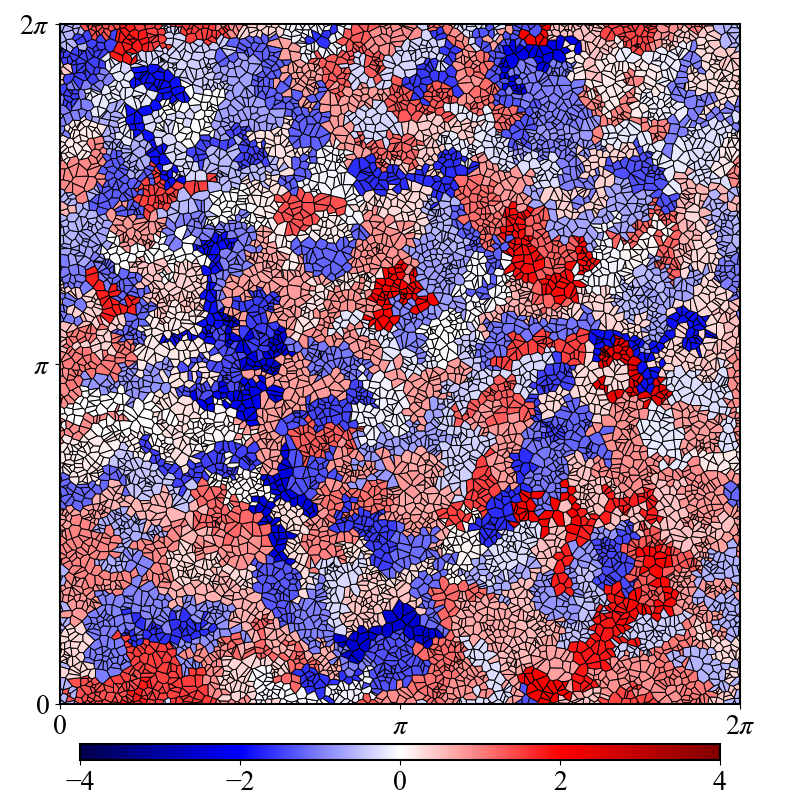}
    \end{minipage}
    \hspace{-5mm}
    \begin{minipage}{0.34\linewidth}
    (h)\\
    \includegraphics[width=\linewidth]{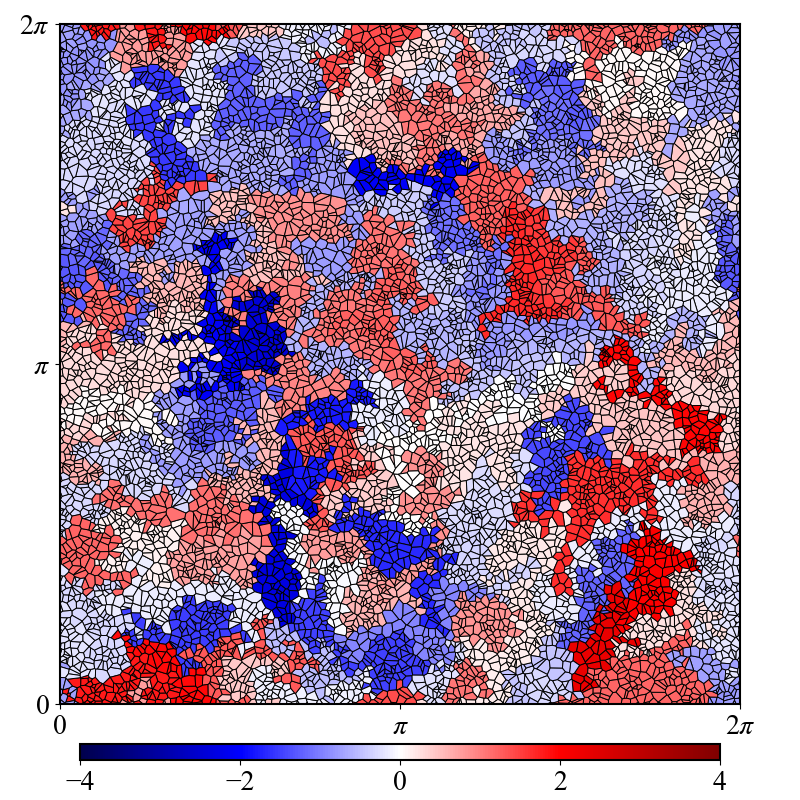}
    \end{minipage}
    \hspace{-5mm}
    \begin{minipage}{0.34\linewidth}
    (i)\\
    \includegraphics[width=\linewidth]{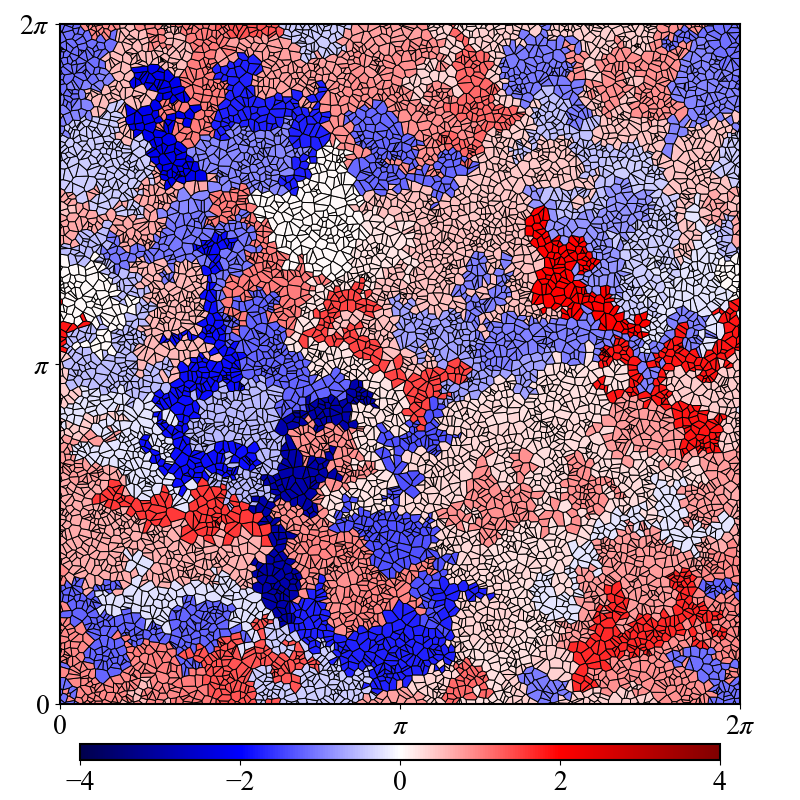}
    \end{minipage}
    \caption{Multiresolution tessellation for a 
    2D uniform random particle distribution for levels (a) 0, (b) 1, (c) 2, (d) 3, (e) 4, (f) 5, (g) 6, (h) 7, and (i) 8. The number of particles is $N_p=10^4$. The tessellation cells are colored with the coarse-grained signals based on the Gaussian random noise.
    }
    \label{fig:MR_2drandom}
\end{figure}

The tessellation cells and the projected signals $\bar{s}^\ell_i$ for each level $\ell$ for the Gaussian noise are visualized in Figure \ref{fig:MR_2drandom}. The number of particles is $N_p=10^4$. The signals are normalized by the standard deviation of the signal at each level.
We can observe that the size of the same color regions increases as the level increases, and this indicates that the tessellations at different levels have different spatial scales.

\begin{figure}
    \centering
    \begin{minipage}{0.49\linewidth}
    (a)\\
    \includegraphics[width=\linewidth]{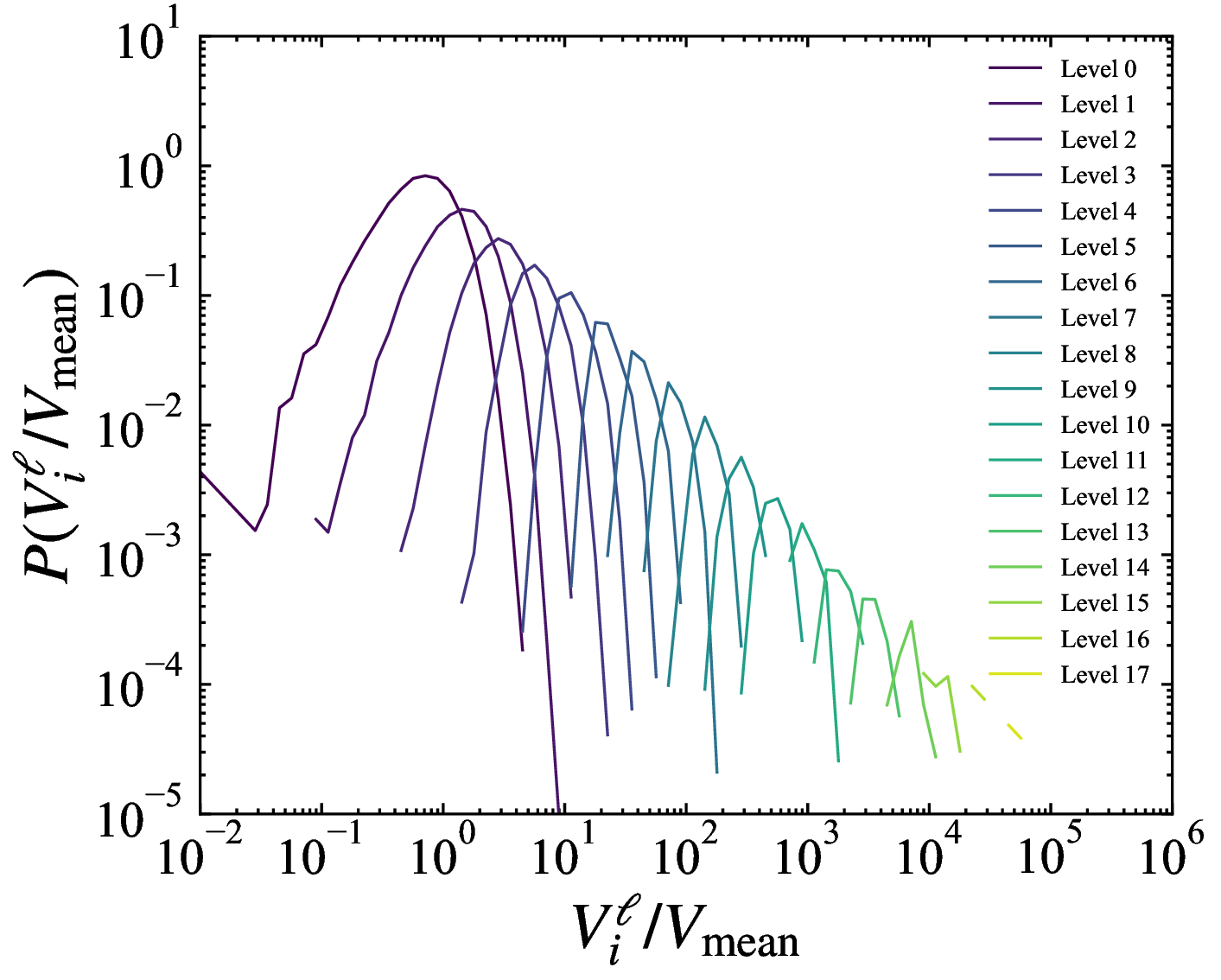}
    \end{minipage}
    \begin{minipage}{0.49\linewidth}
    (b)\\
    \includegraphics[width=\linewidth]{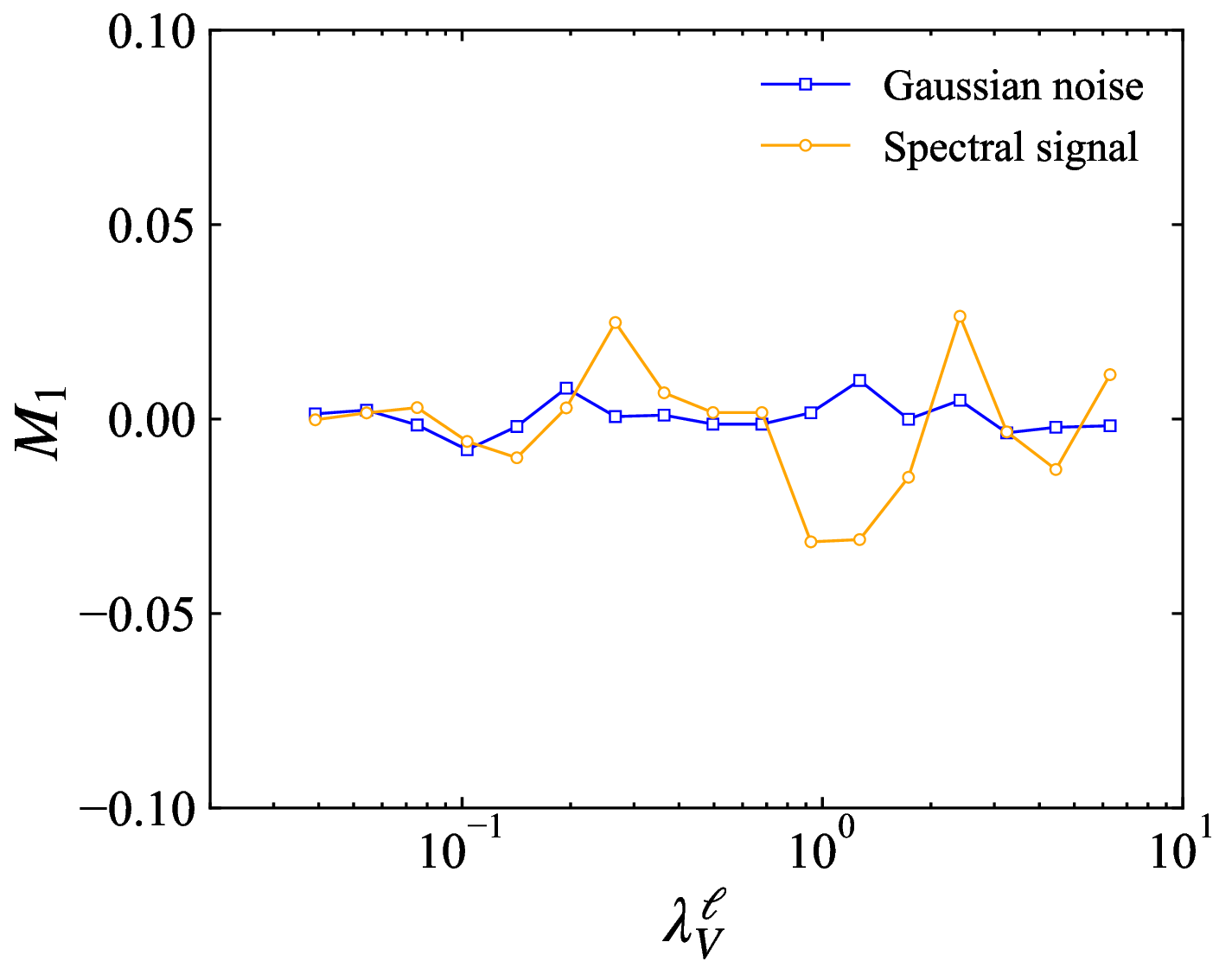}
    \end{minipage} \\
    \begin{minipage}{0.49\linewidth}
    (c)\\
    \includegraphics[width=\linewidth]{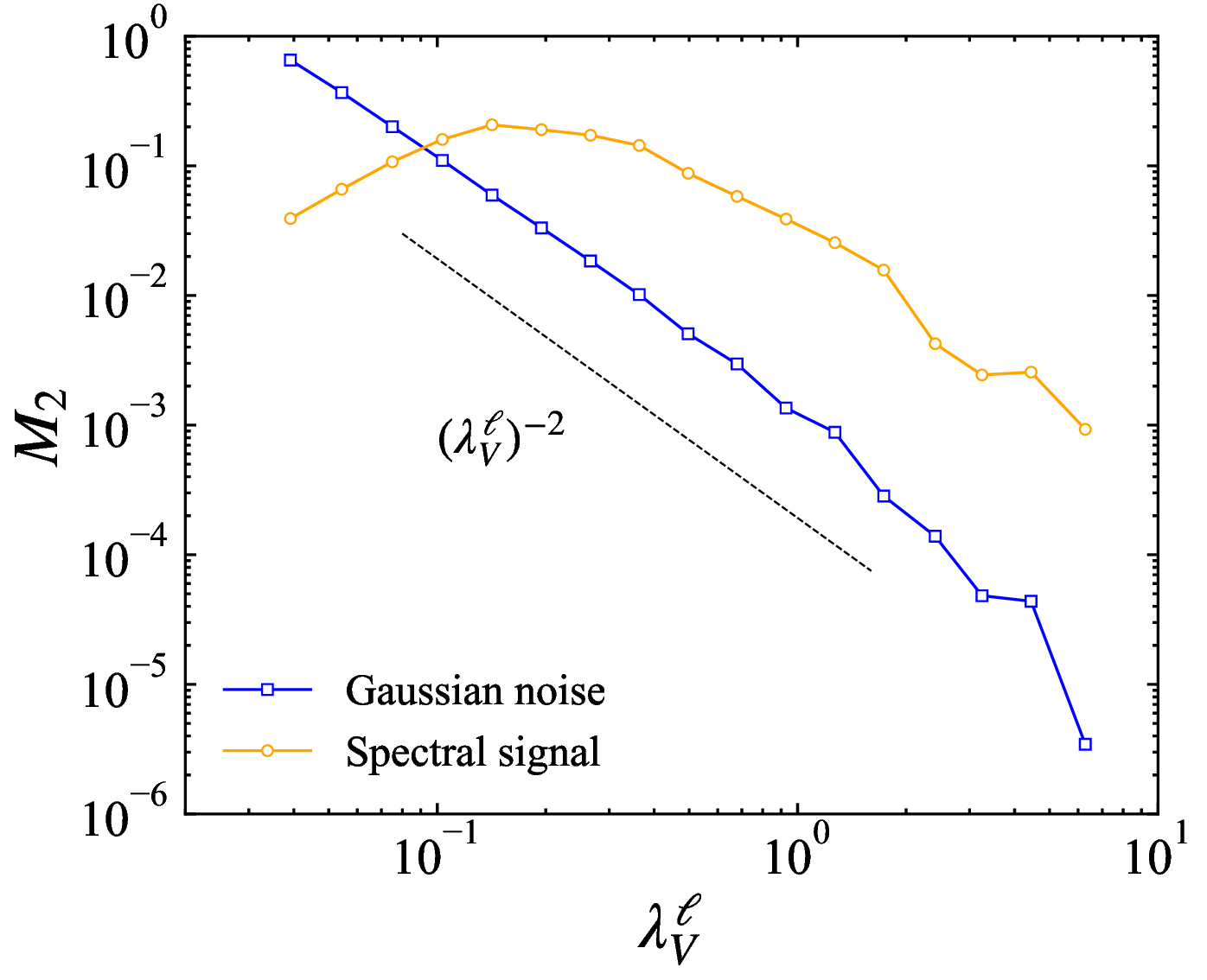}
    \end{minipage}
    \begin{minipage}{0.49\linewidth}
    (d)\\
    \includegraphics[width=\linewidth]{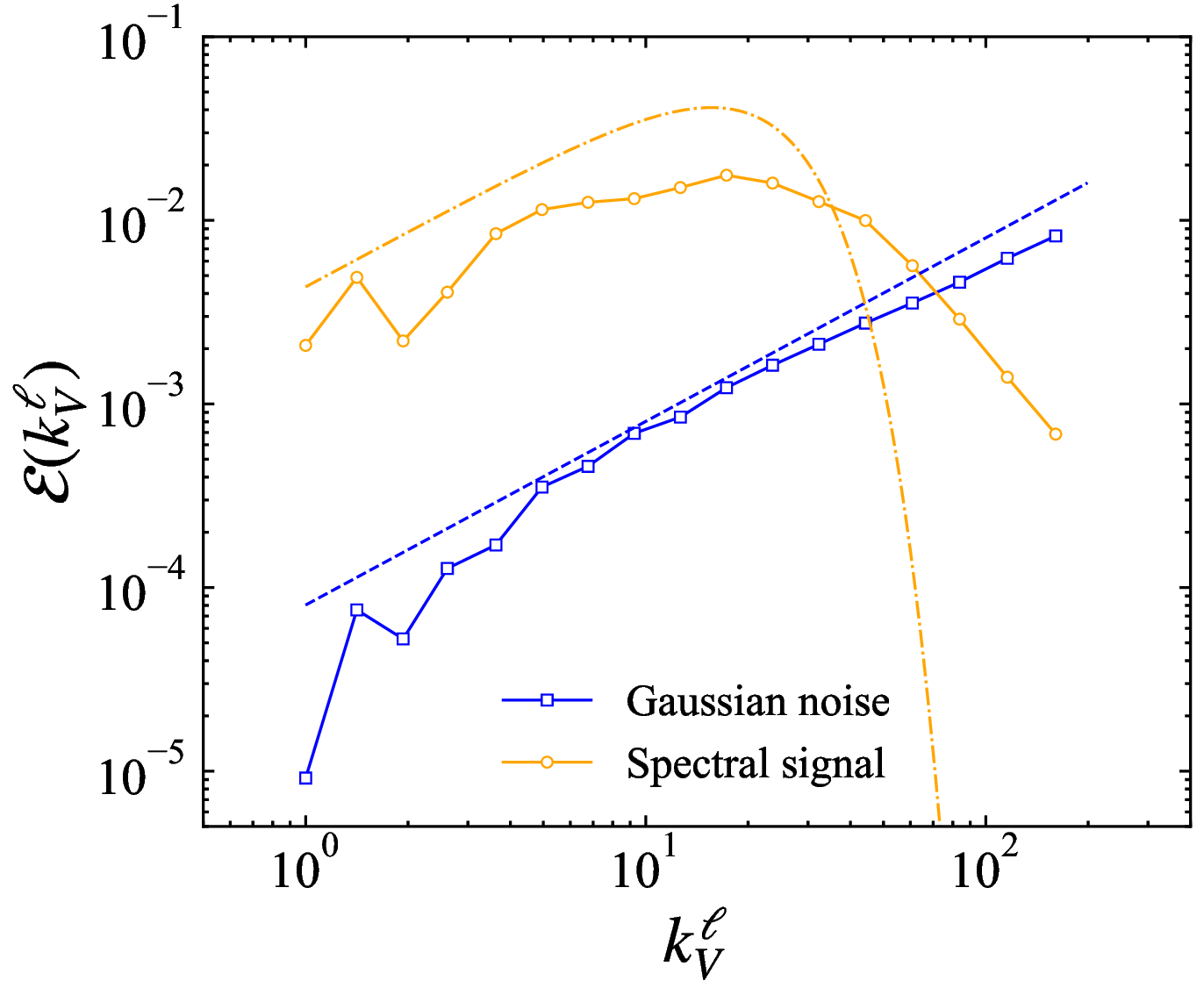}
    \end{minipage}
    \caption{(a) Probability density functions (PDFs) of volume, (b) first and (c) second order moments of detail coefficients, and (d) wavelet spectra for the Gaussian noise and the spectral signal on random particles with    $N_p=10^5$. The dashed and dash-dotted lines show theoretical curves of the Fourier energy spectra for the Gaussian noise and the spectral signal, respectively.}
    \label{fig:statistics_2d}
\end{figure}

Figure \ref{fig:statistics_2d} shows the statistics obtained from the 
2D multiresolution analyses for $N_p=10^5$.
Figure \ref{fig:statistics_2d}(a) presents the probability density function (PDF) of volume $V_i^\ell$ for each level $\ell$. 
The volume is normalized by the mean volume $V_{\rm mean}$. 
As the level increases, the peak of the PDF shifts to larger volume.
The first and second order moments of the detail coefficients are shown in Figures \ref{fig:statistics_2d}(b) and (c), respectively.
The first order moment $M_1$ shows some deviations from zero, and the absolute values are smaller than 6\% of the standard deviation of the signal for both Gaussian noise and spectral signal cases. 
These are stochastic errors of the given signals and they are expected to decrease by taking the ensemble average. 
The second order moment for the Gaussian noise shows the 
scaling close to 
$(\lambda_V^\ell)^{-2}$,
and those for the spectral signal shows a peak around 
$\lambda_V^\ell \approx 10^{-1}$.
%
Figure \ref{fig:statistics_2d}(d) shows the wavelet energy spectrum.
For Gaussian noise, ${\cal E}(k_V^\ell)$ 
is approximately proportional to $k_V^\ell$, 
as expected for the Fourier energy spectrum (dashed line) estimated by $k\sum_{i=0}^{N_p-1}V_i^2/(2\pi)$.
For the spectral signal, ${\cal E}(k_V^\ell)$ 
shows rapid decay for $k_V^\ell \gtrsim k_c$, 
having a peak around $k_V^\ell \sim k_c$. 
These results are consistent with the characteristics of the Fourier energy spectra for these signals. Therefore, we can conclude that the present multiresolution technique can capture the scale-dependence of the signals at particle positions.
The detailed differences between the wavelet and Fourier energy spectra will be discussed in the next subsection.


%

\subsection{Three-dimensional isotropic turbulence}
\label{app3d}
%
\begin{figure}
    \centering
    (a)\includegraphics[width=0.7\linewidth]{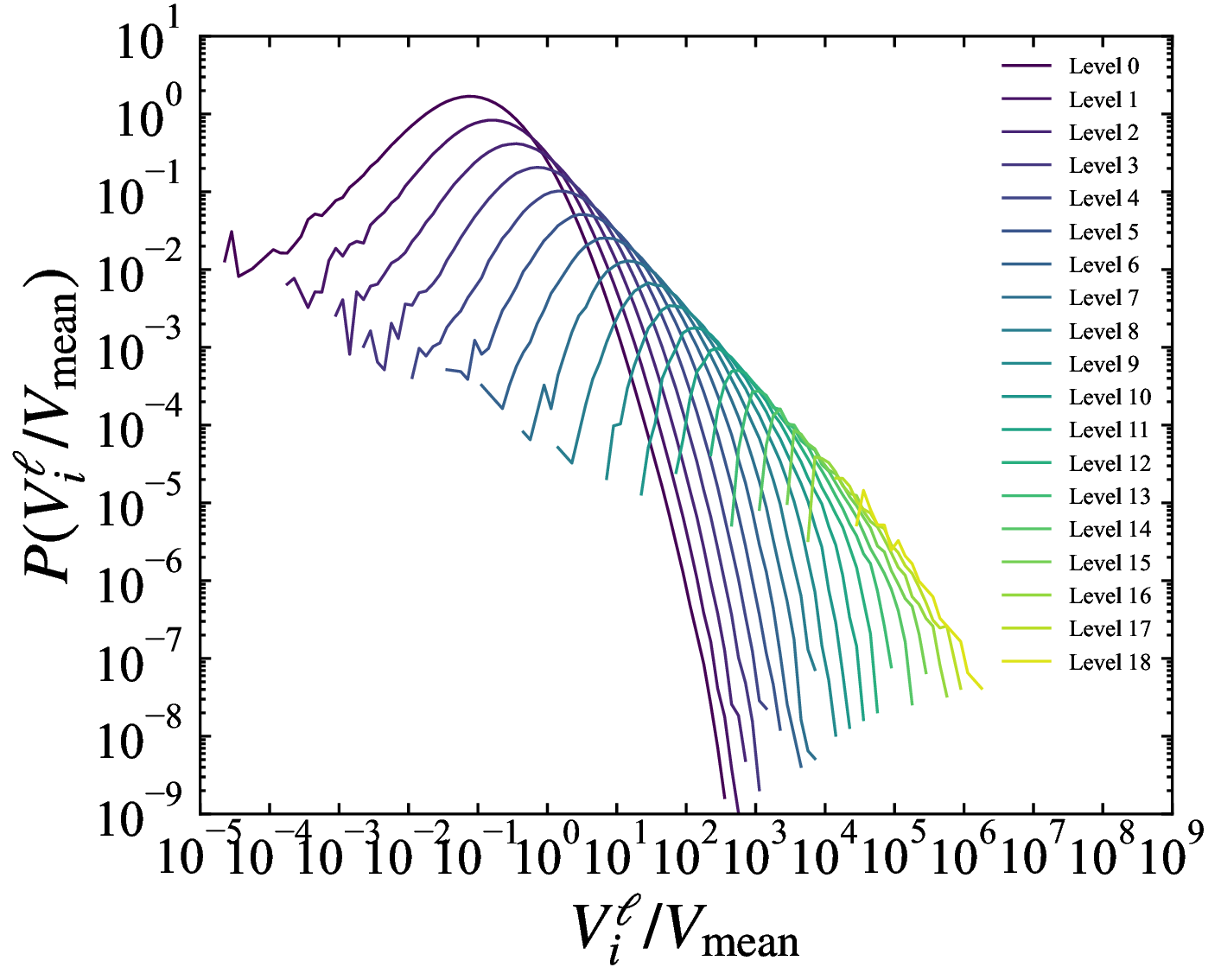} \\
    (b)\includegraphics[width=0.7\linewidth]{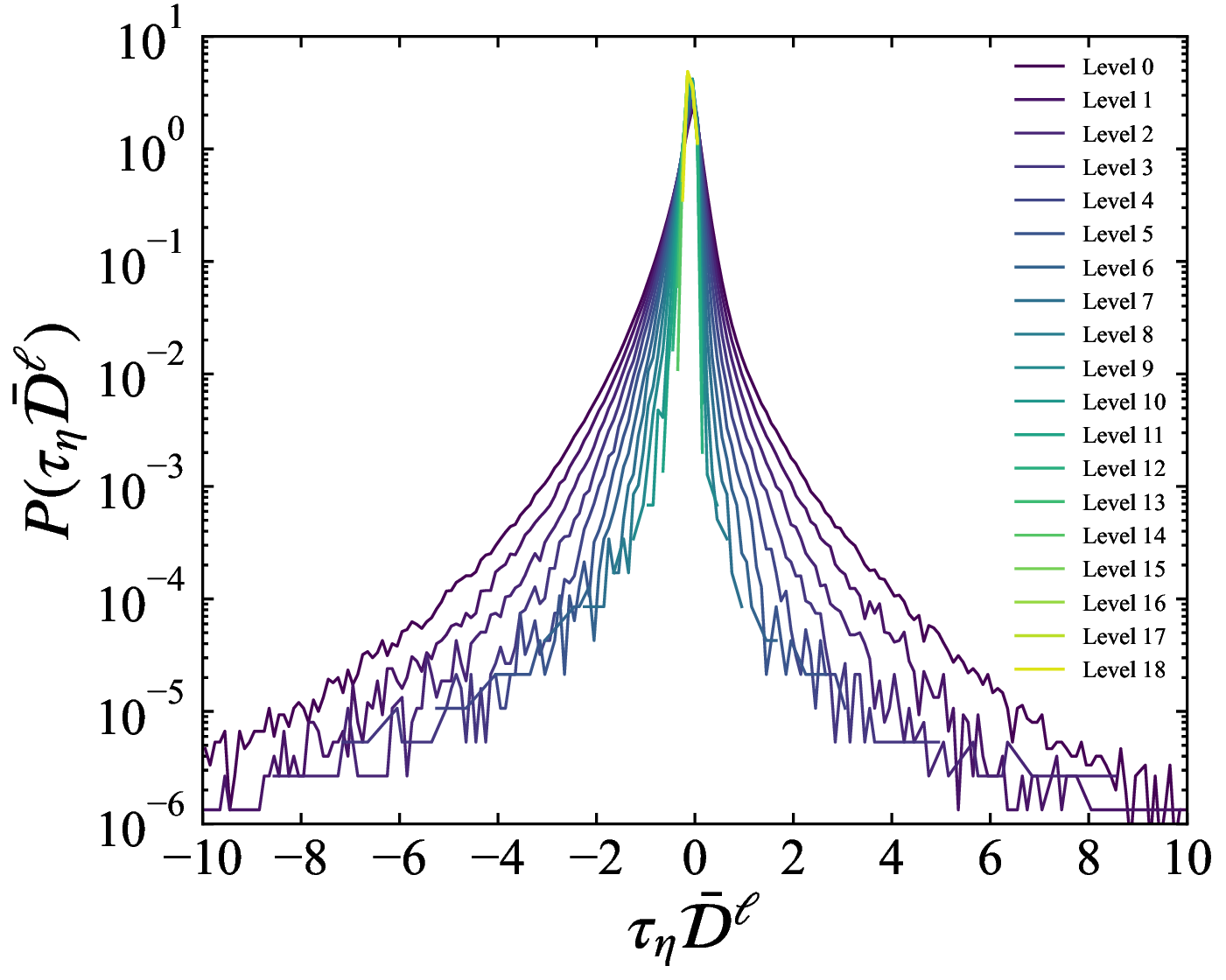}
    \caption{
    PDFs of (a) volume and (b) projected particle velocity divergence for inertial particles in HIT for $Re_\lambda=204$, $St=1$, and $N_p=1.5\times10^7$.
    }
    \label{fig:PDF_vol_div_HIT}
\end{figure}
We apply the proposed multiresolution technique to the particle velocity divergence ${\cal D}$ for inertial particles in 
3D HIT, i.e., $s_i^0 = {\cal D}({\bm x}_{p,i})$. 
We used the particle position and velocity data obtained from the DNS by \cite{Matsuda2014}.
The HIT was computed in the cubic computational domain with edge length $2\pi$. The number of grid points is $512^3$, and the Taylor microscale Reynolds number is $Re_\lambda = 204$. The Stokes number based on the Kolmogorov time $\tau_\eta$ (i.e., $St \equiv \tau_p/\tau_\eta$) is $St=1.0$. The number of particles is $N_p=1.5\times10^7$. The mean particle number density is $\langle n\rangle\eta^3 = 0.030$.
The particle velocity divergence $\cal D$ is calculated using the tessellation technique described in Section~2.

Figure \ref{fig:PDF_vol_div_HIT}(a) shows the PDF of volume $V_i^\ell$ for each level $\ell$ for inertial particles in HIT. 
Similar to the case of 
2D random particles, the peak of the PDF moves to larger volume as the level increases. However, the separation of the scale is not as clear as in the 2D case because the variance of the tessellation cell volume for clustering particles is larger than that for random particles.
Figure \ref{fig:PDF_vol_div_HIT}(b) shows the PDF of the projected values of $\cal D$ for each level $\ell$.
The PDF becomes narrower as the level increases. This confirms that the present multiresolution technique can be used for coarse graining 
the divergence values at discrete particle positions. 

\begin{figure}
    \centering
    \includegraphics[width=0.7\linewidth]{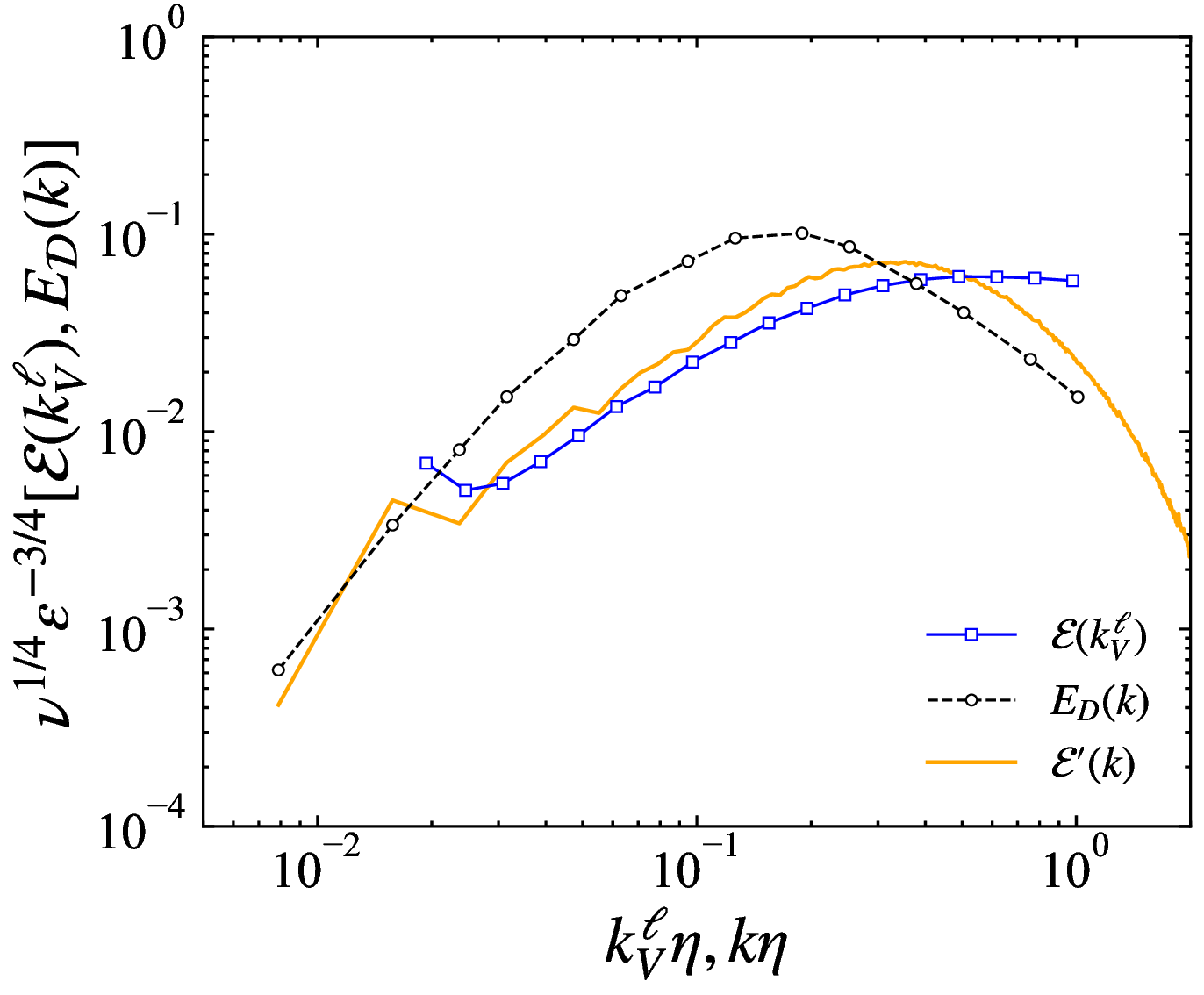} 
    \caption{Wavelet and Fourier energy spectra of the particle velocity divergence $\cal D$ for inertial particles in HIT for $Re_\lambda=204$, $St=1.0$, and $N_p=1.5\times10^7$.
    }
    \label{fig:M2_E2_HIT}
\end{figure}

%
Figure~\ref{fig:M2_E2_HIT} shows the wavelet energy spectrum ${\cal E}(k_V^\ell)$ in comparison with the Fourier spectrum, which is defined as $E_{\cal D}(k)=\sum_{k-1/2\le |{\bm k}|<k+1/2} \widehat{\cal D}({\bm k})\widehat{\cal D}^*({\bm k})$, where $\widehat{\cal D}({\bm k})$ is the Fourier coefficient of ${\cal D}({\bm x})$ and computed by the analytical Fourier transform \citep{Matsuda2014}:
Since the divergence values are defined only at particle positions, the spatial distribution of ${\cal D}({\bm x})$ can be expressed as 
${\cal D}({\bm x}) = \sum_{i=0}^{N_p-1} {\cal D}({\bm x}_{p,i}) \phi^0_i({\bm x})$, where $\phi^0_i({\bm x})$ is the scaling function defined by Eq.~(\ref{eq:phi_0}) in \ref{app1}. This function takes $\phi^0_i({\bm x})=1$ when ${\bm x}$ is in the tessellation cell of index $i$ while $\phi^0_i({\bm x})=0$ otherwise. 
Here, we approximately give the divergence field as ${\cal D}({\bm x}) \approx \sum_{i=0}^{N_p-1} {\cal D}({\bm x}_{p,i}) V_i \delta({\bm x}-{\bm x}_{p,i})$, where $\delta({\bm x})$ is the Dirac delta function, and apply the Fourier transform. The Fourier coefficients are then given by
\begin{equation}
    \widehat{\cal D}({\bm k}) = (2\pi)^{-3}\sum_{i=0}^{N_p-1} {\cal D}({\bm x}_{p,i}) V_i \exp(-\iota{\bm k}\cdot{\bm x}_{p,i})
    \label{eq:Fourier_nD},
\end{equation}
where $\iota = \sqrt{-1}$.
It should be noted that the approximation for the divergence field is valid only when the wavelength is sufficiently larger than the volume scale. This limitation does not significantly affect the comparison here since the wavelet spectrum ${\cal E}(k_V^\ell)$ based on the multiresolution tessellation is defined only for the scales larger than the tessellation volume scale.
We can observe that the scaling of the Fourier spectrum $E_{D}(k)$ changes around $k\eta \approx 0.2$. This scale is close to the peak wavenumber of the Fourier energy spectrum of the number density ($k\eta \approx 0.2$) reported by \cite{Matsuda2014}.
%
The wavelet spectrum $E_{\rm MR}(k_V)$ shows similar values as 
the Fourier spectrum, while the scaling and the peak locations are not exactly the same as the Fourier spectrum: 
The slope of ${\cal E}(k_V^\ell)$ for $k_V^\ell\eta \lesssim 0.1$ is gentler than the slope of $E_{\cal D}(k)$ for $k\eta \lesssim 0.06$.
The peak location of ${\cal E}(k_V^\ell)$ is $k_V^\ell\eta \approx 0.5$, i.e., it is shifted to higher wavenumber.

\begin{figure}
    \centering
    \includegraphics[width=0.7\linewidth]{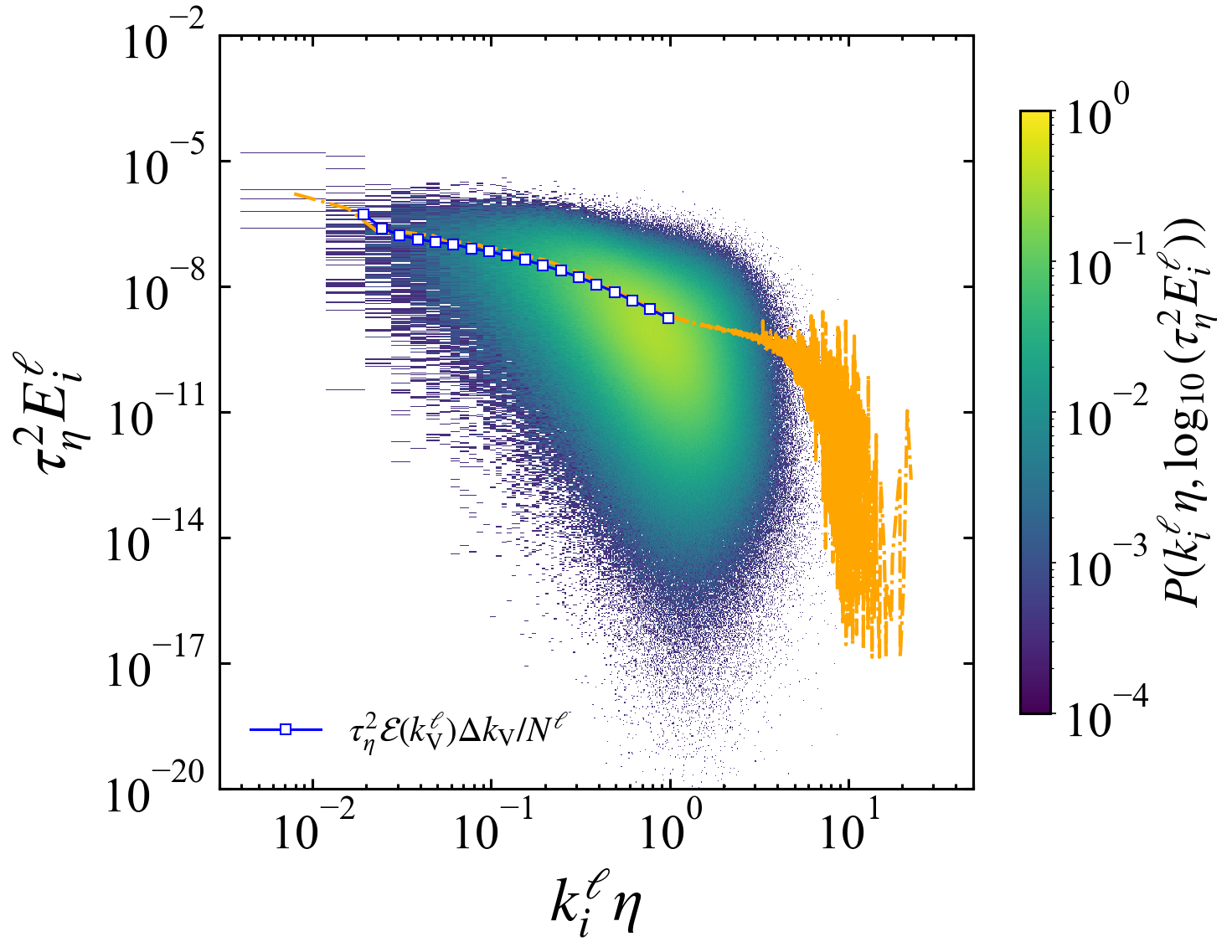} 
    \caption{Joint PDF of the normalized wavenumber $k_i^\ell \eta$ and the logarithmically normalized energy $\tau_\eta^2 E_i^\ell$ of each wavelet coefficient for $\cal D$ in the HIT. 
    The orange dash-dotted line is the conditionally averaged energy $\tau_\eta^2 \langle E^\ell_i \rangle_{k}$ for each wavenumber $k_i^\ell$.
    }
    \label{fig:jointPDF_HIT}
\end{figure}

The differences between the wavelet and Fourier spectra are mainly caused by the spread of the tessellation volumes, as observed in Figure \ref{fig:PDF_vol_div_HIT}, and the effective wavenumber of the wavelets $\psi_i^\ell$. 
Since the present wavelets are constructed based on the tessellation volumes, the scales of wavelets in the same level are also dispersed. 
Figure \ref{fig:jointPDF_HIT} shows the joint PDF 
$P(k_i^\ell \eta,\log_{10}(\tau_\eta^2 E^\ell_i))$ of the wavenumber $k_i^\ell$ and the wavelet energy $E_i^\ell$ for each detail coefficient of $\cal D$ from the HIT data.
The open squares represent the wavelet energy spectrum scaled to represent the mean wavelet energy at each level.
We can observe that the wavelet coefficients are dispersed in wider wavenumber and energy ranges than the wavelet energy spectrum ${\cal E}(k_V^\ell)$. 
It is also observed that the scaled wavelet spectrum ${\cal E}(k_V^\ell)$ is close to the conditionally averaged energy $\tau_\eta^2 \langle E^\ell_i \rangle_{k} = \tau_\eta^2 \int_0^\infty E^\ell_i P(k_i^\ell, E^\ell_i) dE^\ell_i/P(k_i^\ell)$ displayed in the orange dash-dotted line, where $P(k_i^\ell)$ is the PDF of $k_i^\ell$. 
This result suggests that the wavelet energy spectrum well captures the scale dependence of the particle velocity divergence.
To take account of the spread of the wavenumber $k_i^\ell$ and the wavelet energy $E_i^\ell$ for each detail coefficient,
the wavenumber-based wavelet energy spectrum ${\cal E}'(k)$ is plotted in Figure \ref{fig:PDF_vol_div_HIT} by applying the same normalization as ${\cal E}(k_V^\ell)$ and $E_{\cal D}(k)$. The wavenumber-based spectrum ${\cal E}'(k)$ shows a shifted peak similar to the level-based spectrum ${\cal E}(k_V^\ell)$ whereas the slope is steeper than ${\cal E}(k_V^\ell)$.
The difference between the wavenumber-based wavelet energy spectrum ${\cal E}'(k)$ and the Fourier spectrum $E_{\cal D}(k)$ is attributed to the difference in the definitions of the wavenumber.
Identifying the effective wavenumber, such as the centroid wavenumber, for the present wavelets would help completely describing the relationship between the wavelet spectrum and the Fourier spectrum.



\begin{figure}
    \centering
    \begin{minipage}{0.49\linewidth}
    (a)\\
    \includegraphics[width=\linewidth]{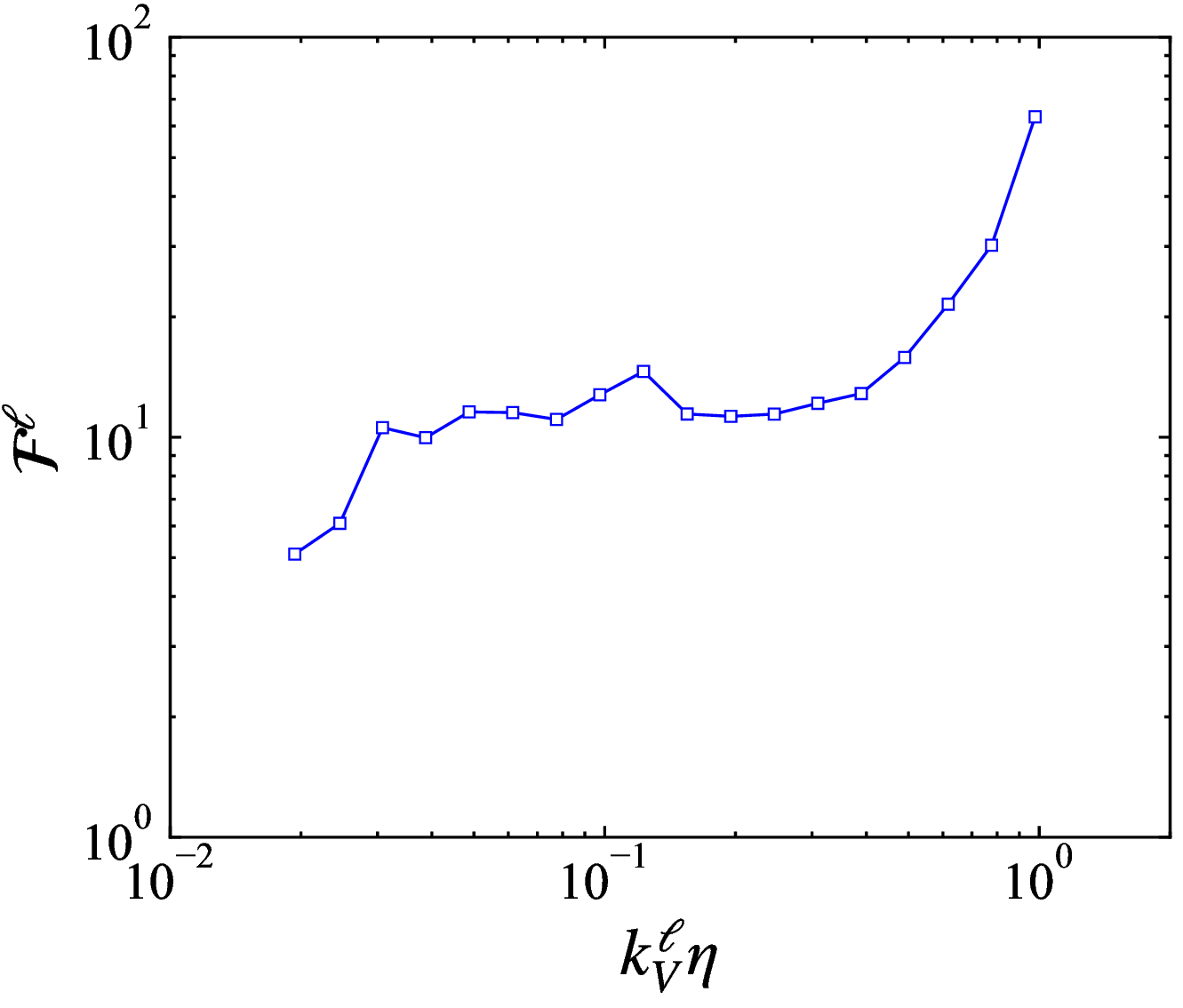}
    \end{minipage}
    \begin{minipage}{0.49\linewidth}
    (b)\\
    \includegraphics[width=\linewidth]{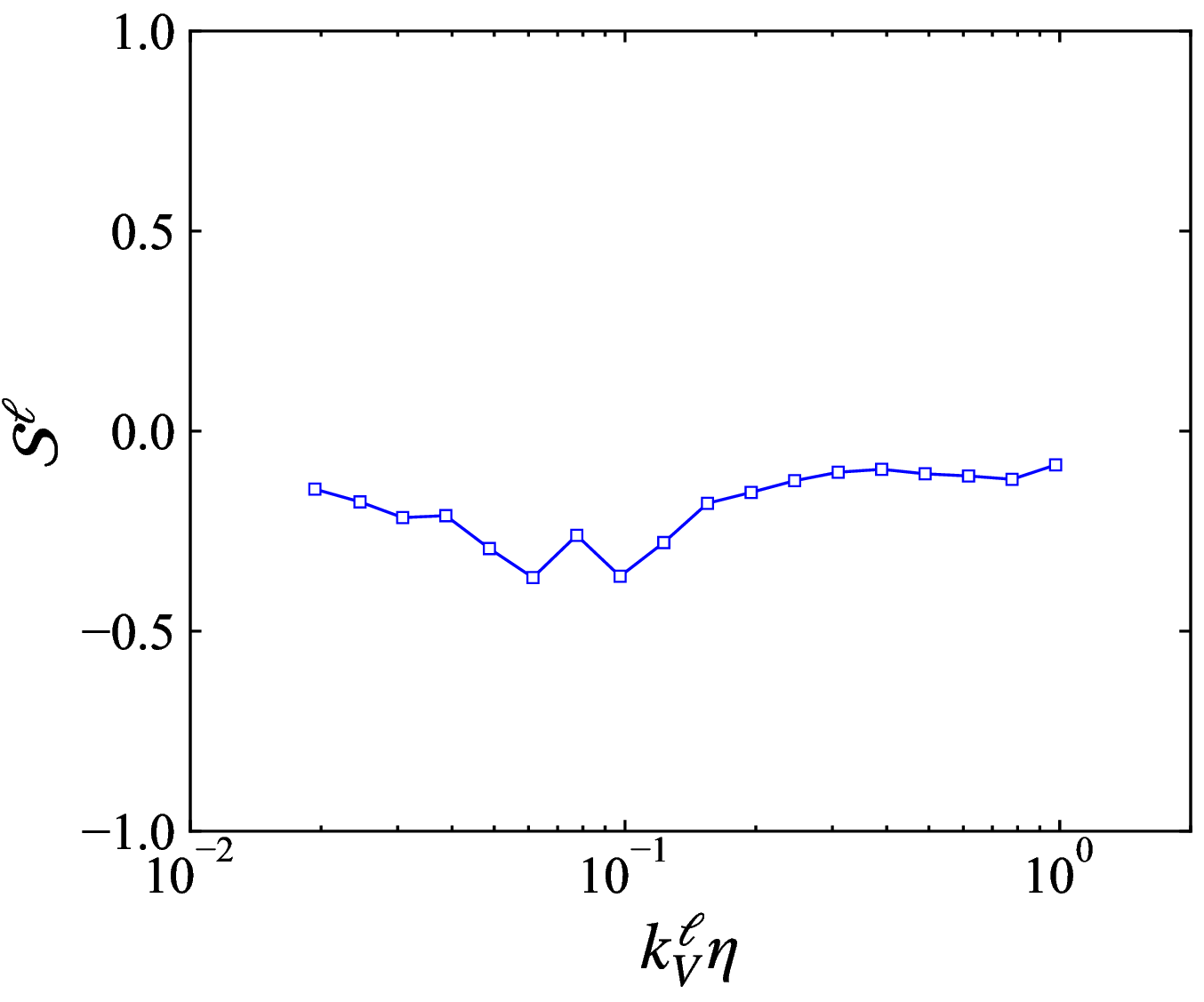}
    \end{minipage}
    \caption{(a) Scale-dependent flatness ${\cal F}^\ell$ and (b) skewness ${\cal S}^\ell$ of the band-pass filtered particle velocity divergence $\check{\cal D}^\ell$ for inertial particles in the HIT.}
    \label{fig:Flatness_Skewness_HIT}
\end{figure}

We further analyze the band-pass filtered field $\check{\cal D}^\ell$ of the particle velocity divergence. Figure \ref{fig:Flatness_Skewness_HIT} shows the scale-dependent flatness ${\cal F}^\ell$ and skewness $S^\ell$ evaluated from $\check{\cal D}^\ell$ for each level $\ell$. 
The flatness for $0.03 \lesssim k_V^\ell\eta \lesssim 0.4$ ($5 \le \ell \le 16$) is nearly constant, ${\cal F}^\ell \approx 10$, whereas that for $k_V^\ell\eta \gtrsim 0.4$ ($1 \le \ell \le 4$) increases with the increase in the wavenumber $k_V^\ell\eta$. Since the flatness for Gaussian probability distribution results in 3, the flatness larger than 3 indicates intermittent distribution of the divergence $\check{\cal D}^\ell$, and the intermittency becomes more significant for $k_V^\ell\eta \gtrsim 0.4$ as the scale becomes smaller. 
This wave number range approximately corresponds to the wavenumbers larger than the peak wavenumber of the wavelet energy spectrum.
The scale dependence of the skewness ${\cal S}^\ell$ is less significant compared with the flatness ${\cal F}^\ell$. The skewness is negative and close to zero for all the scales examined here. Negative skewness indicates higher probability of significant cluster formation than void formation. This suggests that cluster formation is relatively pronounced for all the scales.

\begin{figure}
    \centering
    \begin{minipage}{0.50\linewidth}
    (a)\\
    \includegraphics[width=\linewidth]{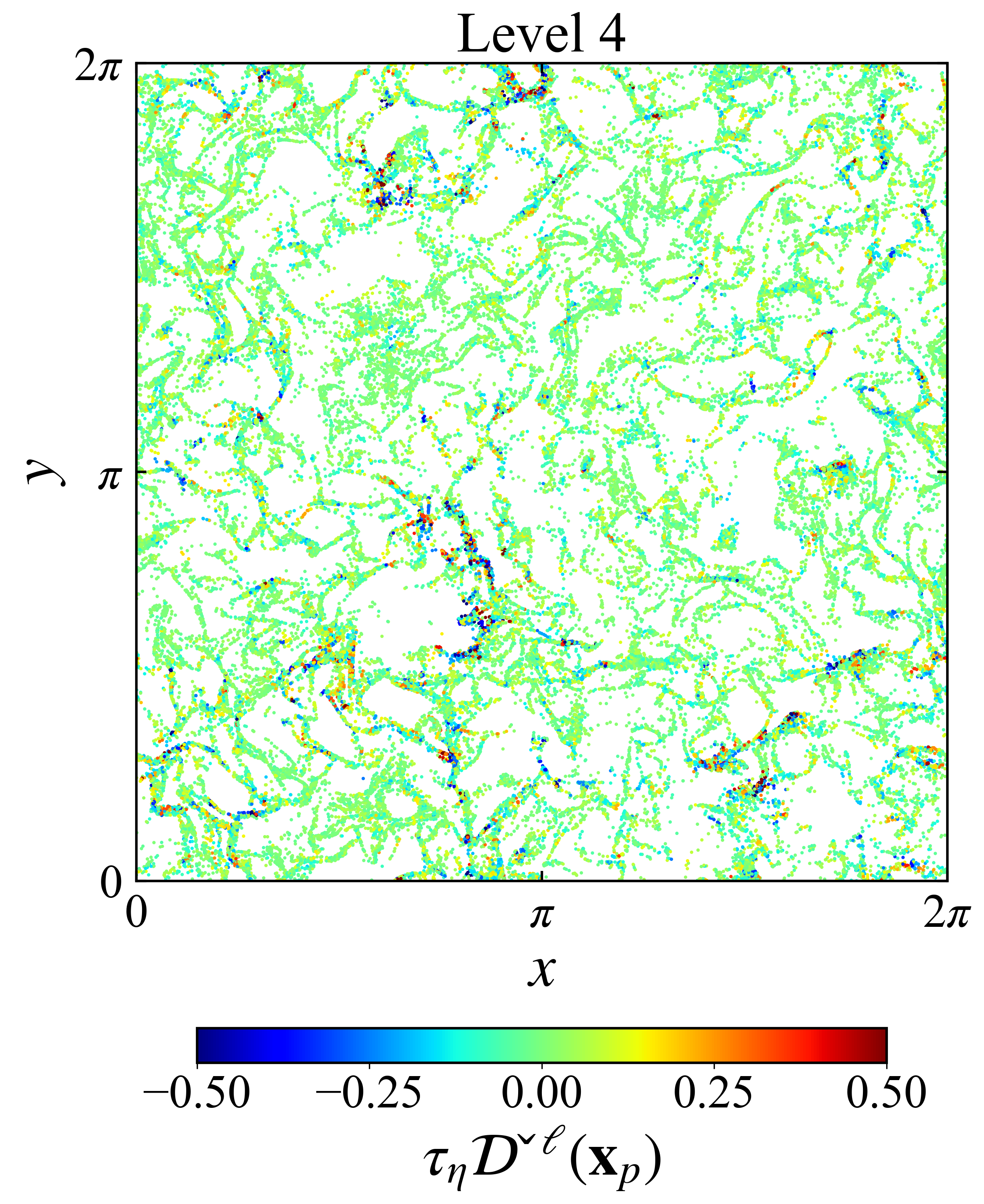}
    \end{minipage}
    \hspace{-3mm}
    \begin{minipage}{0.50\linewidth}
    (b)\\
    \includegraphics[width=\linewidth]{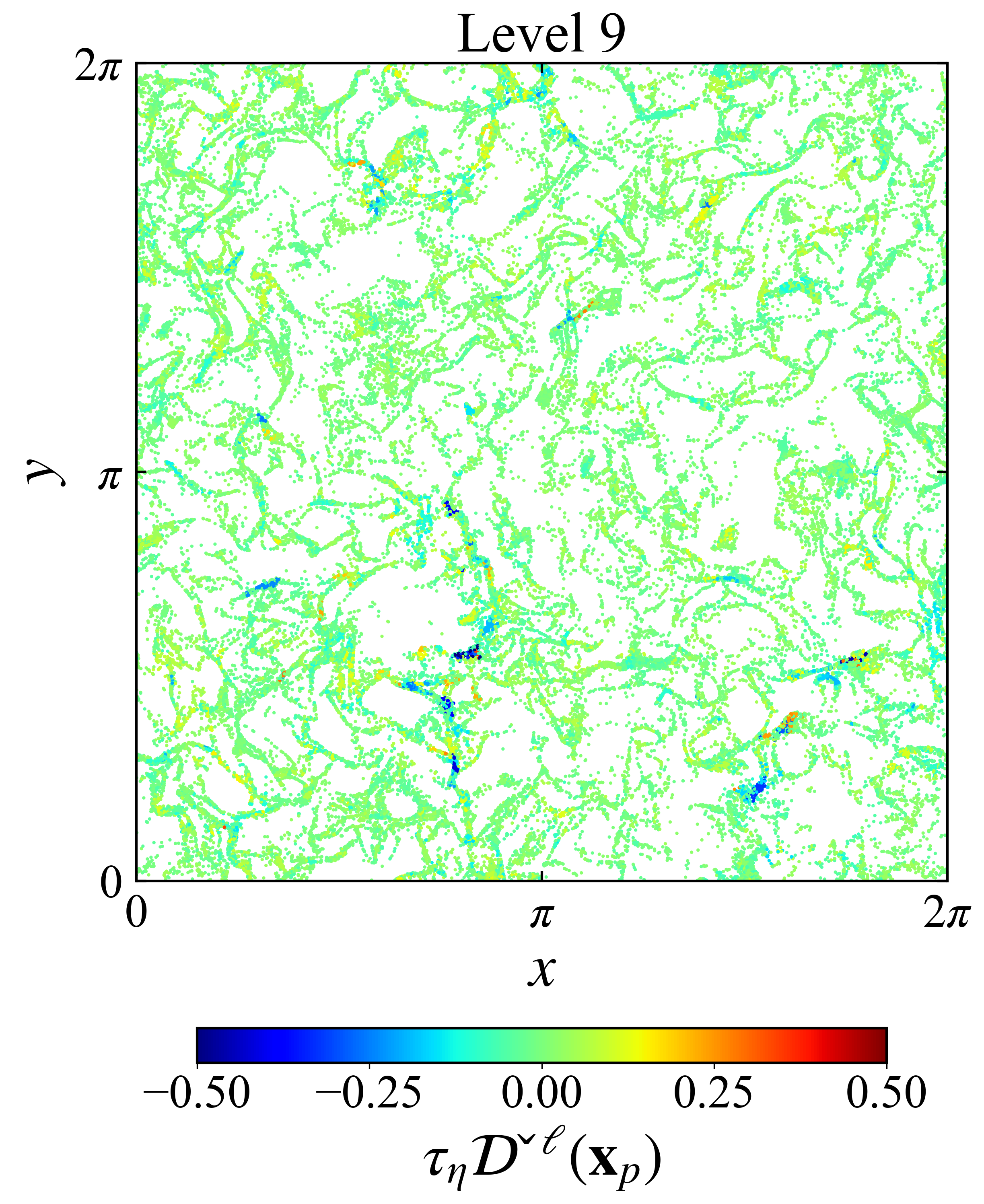}
    \end{minipage}
    \caption{Comparison of band-pass filtered particle velocity divergence $\check{\cal D}^\ell$ at $\ell=4$ and $\ell=9$.
    The particles in the range of $0 \le z \le 4\eta$ are 
    plotted with the color depending on the $\check{\cal D}^\ell$ value for each particle.}
    \label{fig:Bandpass_HIT}
\end{figure}

Figure \ref{fig:Bandpass_HIT} displays the spatial distributions of the band-pass filtered values $\check{\cal D}^\ell$ of the divergence at levels $\ell=4$ and 9. The particles located in the range of $0 \le z \le 4\eta$ are plotted with color of the band-pass filtered divergence value.
Level 4 corresponds to the peak location $k_V^\ell \eta \approx 0.5$ in the wavelet energy spectrum ${\cal E}(k_V^\ell)$, and level 9 is $k_V^\ell \eta \approx 0.15$.
At level 4, the detail information represents the divergence differences inside clusters, whereas at level 9 it represents the difference between clusters. 
This result confirms that the present technique can be used to decompose the particle velocity divergence into detail information at different scales. 
The significant divergence differences inside clusters observed at level 4 suggest the regions where the particle velocity is multi-valued, i.e., particle trajectories can cross, and these are referred to as `caustics' \cite{Wilkinson2005caustics,Meibohm2023,Meibohm2024}. The caustics appear due to the deviation of inertial particle trajectory from the fluid particle trajectory. In the caustic regions, particles can be adjacent to particles with significantly different path histories, and the divergence computation on the Delaunay graph can result in coexistence of significant positive and negative divergence values. Therefore, the significant divergence values caused by the caustics are less relevant to cluster formation. 

We then apply the multiresolution analysis to filter out the significant divergence differences in clusters, aiming to eliminate the effect of caustics. 
We can consider that wavelet coefficients of $\tau_p|d^\ell_i| > {\cal O}(1)$ are caused by the caustics because the inverse of the particle velocity divergence $\cal D$ is the time scale of vanishing or doubling the tessellation cell volume, and the particle dynamics would vary only weakly during the order of the the particle relaxation time $\tau_p$.
Therefore, the filter $f^\ell_i$ ($\ell = 1, \cdots, L$, and $i = 0, \cdots, N^\ell-1$) is defined by $f^\ell_i = 1$ for $i$ and $\ell$ with $\tau_p|d^\ell_i| \le 0.3$ and $f^\ell_i = 0$ otherwise. The filtered divergence ${\cal D}_f$ is obtained by applying the inverse wavelet transform to $f^\ell_i d^\ell_i$ on the multiresolution graphs.

\begin{figure}
    \centering
    \begin{minipage}{0.50\linewidth}
    (a)\\
    \includegraphics[width=\linewidth]{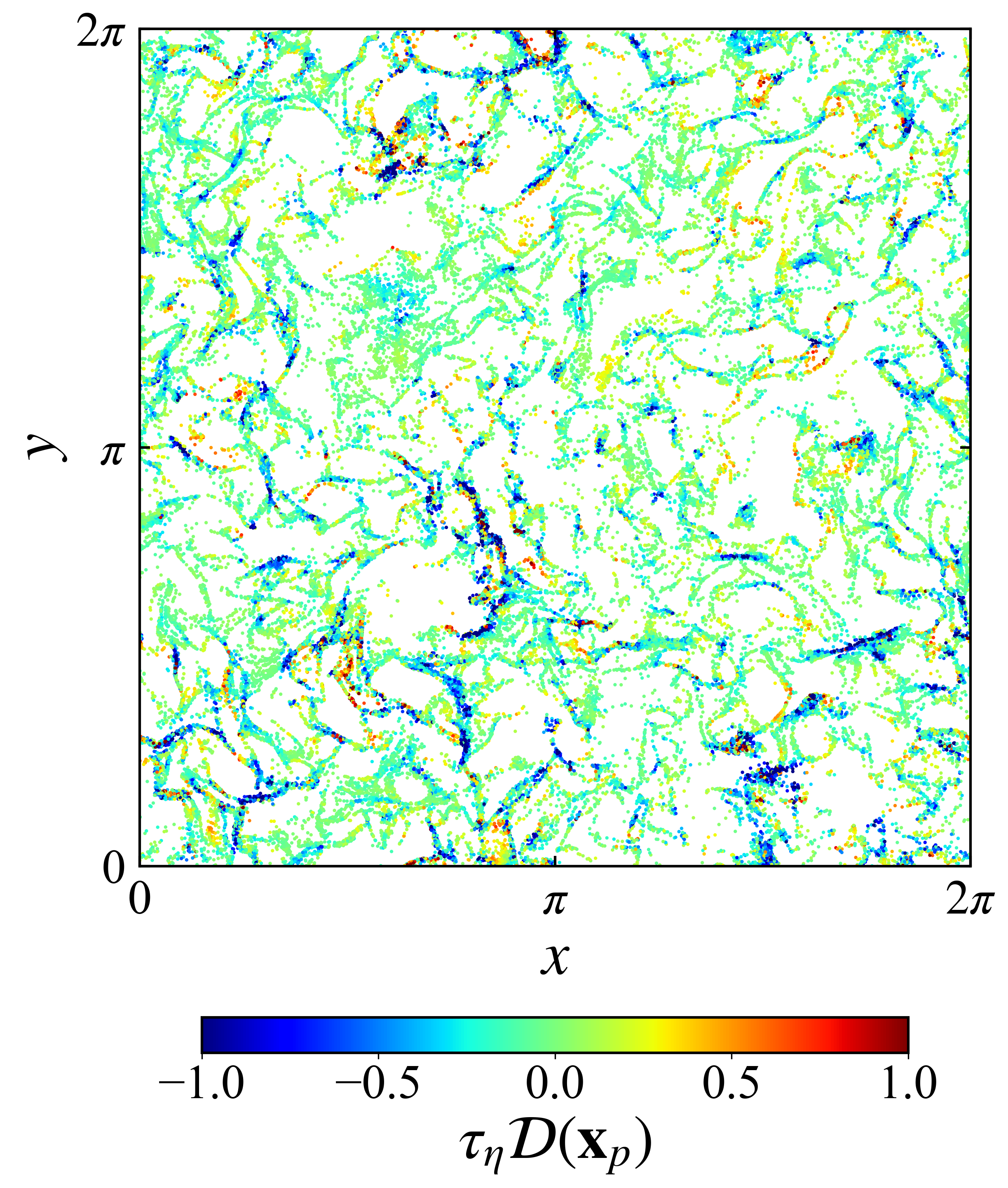}
    \end{minipage}
    \hspace{-3mm}
    \begin{minipage}{0.50\linewidth}
    (b)\\
    \includegraphics[width=\linewidth]{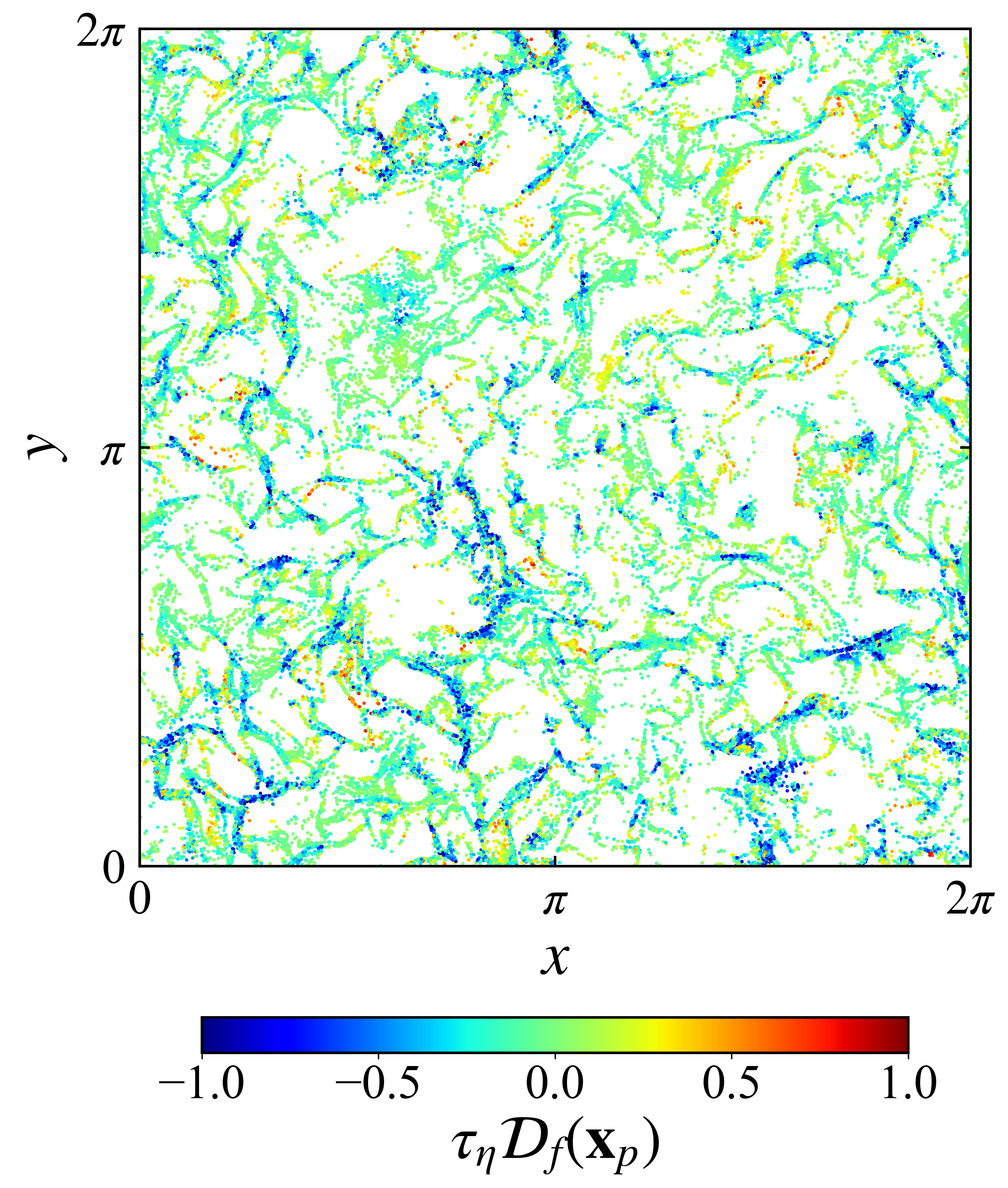}
    \end{minipage} \\
    \begin{minipage}{0.50\linewidth}
    (c)\\
    \includegraphics[width=\linewidth]{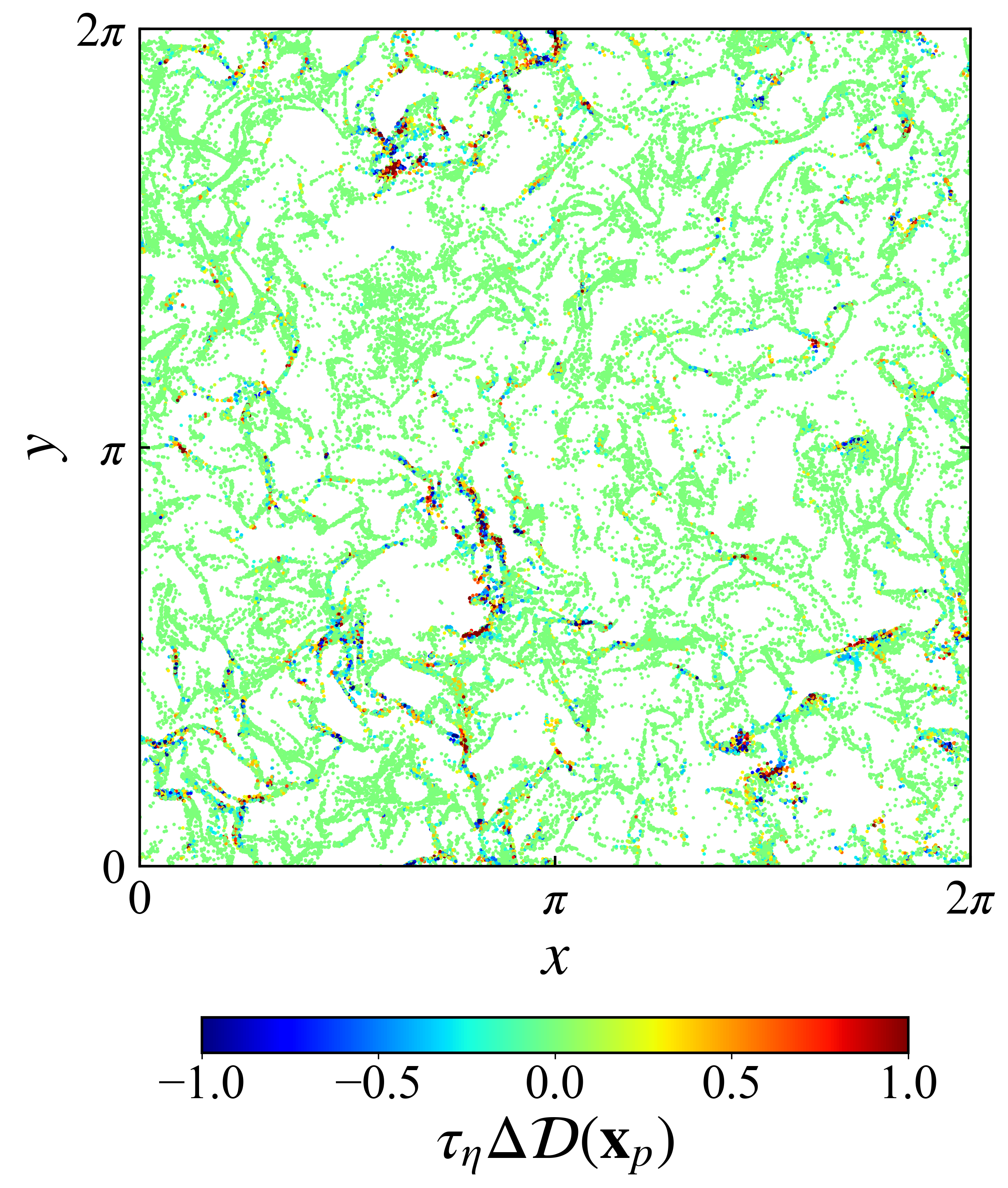}
    \end{minipage}
    \hspace{-3mm}
    \begin{minipage}{0.50\linewidth}
    (d)\\
    \includegraphics[width=\linewidth]{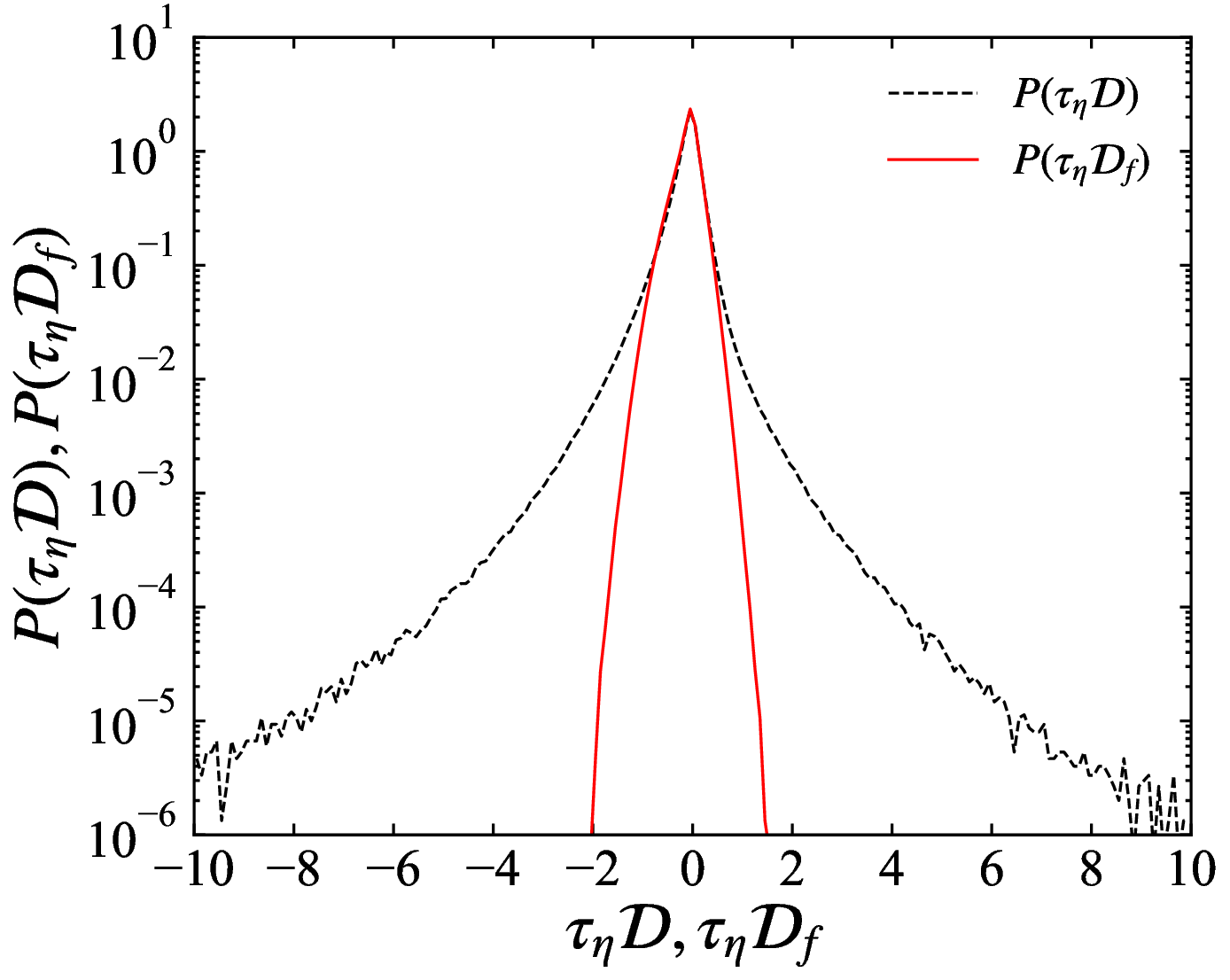}
    \end{minipage}
    \caption{Spatial distributions of (a) original particle velocity divergence ${\cal D}$, (b) filtered divergence ${\cal D}_f$, and (c) the difference $\Delta{\cal D} = {\cal D}_f - {\cal D}$, and (d) the PDF of the original and filtered divergence. 
    The particles in the range of $0 \le z \le 4\eta$ are 
    plotted in  (a), (b), and (c).}
    \label{fig:filtering}
\end{figure}

Figure \ref{fig:filtering} displays the divergence values of (a) the original and (b) filtered data and (c) their difference. 
In figure \ref{fig:filtering}(a), we can observe that significantly large positive and negative divergence values coexist in the same clusters, suggesting the effect of caustics. By applying the filtering to the divergence data, the filtered divergence in figure \ref{fig:filtering}(b) does not show such coexistence of large positive and negative values in clusters. The difference in figure \ref{fig:filtering}(c) confirms that the removed divergence information has 
only significant values confined in clusters, while the moderate divergence distribution remains in the filtered divergence data. 
The PDF of the original and filtered divergence data are plotted in figure \ref{fig:filtering}(d). The present wavelet-based filtering removes the tails on both positive and negative sides, and the filtered divergence values are confined in $\tau_\eta |{\cal D}| \lesssim 2$. The PDF shape of the filtered divergence around $\tau_\eta |{\cal D}| \lesssim 1$ remains similar to that of the original divergence. 
These results indicate that the present multiresolution analysis can be used to decompose the divergence field on particle cloud into the divergence of cluster and void formations and divergence caused by multi-valued particle velocity field.

\section{Conclusions}
\label{conclusion}

We have developed a multiresolution technique on tessellation graphs to evaluate multiscale statistics of field data on particle clouds, aiming to analyze particle velocity divergence defined at discrete particle positions as the first application. 
The Delaunay tessellation was utilized to define the graph, which consists of particles as vertices connected by the Delaunay edges. 
The multiresolution graphs are constructed by a sequential coarsening approach, which starts from the finest scale of level 0, iterating merger of adjacent vertex pairs.
The projection and prediction operators for the multiresolution analysis, i.e., the wavelet analysis on graph, have been proposed being consistent with the graph coarsening procedure.
To consider the physical space on the graph, we have introduced the volume preserving property based on the volumes of Voronoi cells to the graph construction and the multiresolution operators.

The proposed multiresolution technique has been applied to 2D random particle distributions for two test signals, a Gaussian noise and a spectral signal, 
and 
then to a 3D inertial particle distribution with the particle velocity divergence in HIT.
%
In the 2D test cases, we showed that the size of the volumes defined on the multiresolution graphs increases as the level increases, confirming that the tessellations at coarser levels have larger spatial scales. The characteristics of the wavelet energy spectra for the Gaussian noise and the spectral signal are consistent with those of the Fourier spectra, revealing that the present technique can capture the scale-dependence of the signals at particle positions.
%
In the application to the 3D data, we have analyzed the particle velocity divergence at particle positions, which was computed by applying the tessellation-based technique \cite{MaurelOujia2024JCP} to inertial particle data obtained from the DNS from particle-laden HIT.
The PDFs of the volumes and the projected (low-pass filtered) particle velocity divergence show that similar to 2D cases, the spatial scales becomes larger as the level increases, while the spatial scales spread relatively wider than the 2D random particle case, and the present multiresolution technique captures more coarse grained values as the level increases.
%

The wavelet energy spectrum for the particle velocity divergence in the 3D HIT has been compared with the Fourier spectrum of the same divergence. In overall, the wavelet spectrum shows characteristics similar to those of the Fourier spectrum, whereas the wavelet spectrum has a peak shifted to higher wavenumber and a gentler slope compared with the Fourier spectrum. The statistic analyses of the wavenumber and the wavelet energy of each wavelet revealed that the difference between the wavelet and Fourier spectra are mainly due to the spread of spatial scales of wavelets and the evaluation of the wavenumber for the wavelets.

The higher-order statistics, i.e., flatness and skewness, of the band-pass filtered values for the particle velocity divergence showed the scale dependence of the intermittency and the pronounced cluster formation. The clear scale dependence was observed particularly in the flatness, which indicates significant intermittency for high wavenumbers, which approximately correspond to the peak wavenumber in the wavelet energy spectrum and higher wavenumbers. The visualization of the band-pass filtered values indicates that the values at the peak wavenumber represents the divergence difference inside clusters, whereas the values at a smaller wavenumber shows the divergence difference between clusters. The significant divergence observed intermittently inside clusters suggests the effect of caustics, where the particle velocity field is multi-valued and inertial particle trajectories can cross.

To separate the effect of the caustics from the divergence values that indicate cluster formation, we have applied the filtering to the wavelet coefficients of significant divergence differences. The criterion to filter out the wavelet coefficients was 
chosen based on the time scale of the particle clustering dynamics. The results showed that the filtering removes 
significant divergence difference inside clusters, whereas moderate divergence distribution sustained after the filtering. Therefore, the proposed multiresolution techniques can also be used for filtering that considers the scales and the spatial structures of particle cloud dynamics.



The wavelet approach allows us 
to decompose coherent and incoherent signals, i.e., denoising, and has been used 
to extract coherent clusters of inertial particles in turbulent flows \citep{bassenne2017extraction}. 
Hence, the proposed technique for particle cloud data can be used to extract coherent structures of particle velocity divergence in particle-laden turbulence without projecting 
the data onto Cartesian grids.
Especially for large ensembles of particles and in higher dimension this yields a significant advantage in terms of computational efficiency.
As demonstrated in this paper, second order statistics such as 
wavelet energy spectra can be computed much more efficiently than classical Fourier spectra on unstructured data using the `slow' Fourier transform \cite{Matsuda2014}.
%
We have also 
illustrated
that higher-order scale-dependent statistics can be computed, e.g., scale-dependent flatness and skewness \citep{matsuda2021scale}, based on particle cloud data.
These statistics could be important to understand not only the multiscale and intermittent behaviors of particles in 
high Reynolds number turbulence, but also in many other fields, such as in astrophysics, schools of fish, flocks of birds, and crowd dynamics.
%
%
Exploring this new multiresolution avenue 
for particle-laden turbulence and in other fields is left for future work.
In particular, application to the analysis of different types of particle clustering in particle-laden turbulent flow is a part of our current investigations \cite{West2024IJMF,MaurelOujia2025JFM}.

The current wavelet representation on graphs can be further improved by using the (dual) lifting scheme \cite{sweldens1996wavelets, sweldens1998lifting} and hence the number of vanishing moments of the primary and/or dual wavelets can be further increased.
This leads to a better reproduction of polynomials, an increased scale selectivity, and a sparser multiresolution representation of the data.
The procedure to construct coarser graphs could also be important for improving the volume spread for different levels. Different techniques, such as use of edge weights to determine the collapse edge \cite{Dhillon2007}, should be examined for future works.

\section*{Acknowledgements}
We thankfully acknowledge Kazuki Maeda at Purdue University, Suhas S. Jain at Georgia Institute of Technology, and Jacob R. West at Stanford University for the fruitful discussions during the CTR summer program.
This work was initiated and supported by the Center for Turbulence Research (CTR) during the 2022 Summer Program. We acknowledge financial support from the Japan Society for the Promotion of Science (JSPS) KAKENHI [grant numbers JP20K04298, JP23K03686] (K. Matsuda) and the Agence Nationale de la Recherche (ANR) [grant number ANR-20-CE46-0010-01] (T. Maurel--Oujia, K. Schneider). 
The multiresolution analyses for particle-laden turbulent flow data were performed using the Earth Simulator supercomputer system of JAMSTEC.
Centre de Calcul Intensif d’Aix-Marseille is acknowledged for granting access to its high performance computing resources. 

\appendix
\section{Discrete cell average multiresolution and biorthogonal wavelets}
\label{app1}

The connection between the discrete multiresolution analysis and biorthogonal wavelets in the continuous setting is described, respectively for regular grids in one space dimension and then for graphs in two and three dimensions. To this end, we consider the Hilbert spaces $L^2(\Omega)$ where $\Omega \subset \mathbb{R}^n$ ($n= 1, 2$ and $3$) with the inner product 
$\langle f, g \rangle = \int_{\Omega} f({\bm x}) \, g({\bm x}) d{\bm x}$ 
for $f, g \in L^2(\Omega)$ and the induced norm $\Vert f \Vert_2 = \langle f, f \rangle^{1/2}$. Note that the $L^1$ norm is defined by 
$\Vert f \Vert_1 = \int_\Omega | f({\bm x})| d{\bm x}$ 
and the $L^{\infty}$ norm by 
$\Vert f \Vert_{\infty} = \sup_{{\bm x} \in \Omega} |f({\bm x})|$.

\subsection{Case of regular grids}

The above described discrete multiresolution approach on regular grids is directly related to continuous biorthogonal wavelets, with a pair of scaling functions ($\phi, \widetilde \phi$) and a pair of wavelets ($\psi, \widetilde \psi$) which is important to understand the mathematical properties.
To make the connection we need the projection (coarsening) and the detail coefficients, but only implicitly the prediction. Then we have the two filter pairs that define the biorthogonal wavelets.
In one dimension, the domain $\Omega \in \mathbb{R}$ corresponds to an interval and the cells at level $\ell$, $\Omega^\ell_i$ correspond to subintervals with 
$(\Omega^\ell_i)^\circ \cap (\Omega^\ell_j)^\circ = \emptyset$ 
for $i \ne j$, where $^\circ$ denotes the interior of the cell. The length of each subinterval $\Omega^\ell_i$ is denoted by $V^\ell_i$ ($\propto 2^\ell$) which in the case of a regular grid does not depend on $i$.
The cell average value at level $\ell$ of a signal $s$ is then simply given by $\overline{s}^\ell_i = (1/V^\ell_i) \int_{\Omega^\ell_i} \, s(x) dx$ which can be rewritten as $\overline{s}^\ell_i = \langle s , \widetilde \phi^\ell_i \rangle$ where the dual scaling function $\widetilde \phi^\ell_i$ is the indicator function of the cell $\Omega^\ell_i$ divided by its volume. Thus, by construction, we have $\Vert \widetilde \phi^\ell_i \Vert_1 = 1$.

The calculation of the detail coefficients 
$d^{\ell}_i$ 
by subtracting the predicted value 
$\widehat s^{\ell-1}_i$ 
from the exact value 
$\overline{s}^{\ell-1}_i$ 
can be rewritten in terms of a projection onto the dual wavelet $\widetilde \psi^\ell_i$, i.e., 
$d^{\ell}_i = \langle s, \widetilde \psi^{\ell}_i \rangle$ 
where 
$\widetilde \psi^\ell_i = V^{\ell-1}_{2i} / V^{\ell}_i \left( \widetilde{\phi}^{\ell-1}_{2i+1} - \widetilde{\phi}^{\ell-1}_{2i} \right)$ 
in the case of the piecewise constant prediction.
The primary scaling function $\phi^\ell_i$ corresponds in the present context to a piecewise constant function and is therefore only the indicator function of the cell $\Omega^{\ell}_i$, without any scaling, and hence $\Vert \phi^\ell_i \Vert_{\infty} = 1$. The primary wavelet $\psi^\ell_i$ corresponds to the difference of two indicator functions at the coarser level, i.e., $\psi^\ell_i = \phi^{\ell-1}_{2i+1} - \phi^{\ell-1}_{2i}$ implies $\Vert \psi^\ell_i \Vert_{\infty} = 1$.

The biorthogonal wavelet expansion of the signal $s \in L^2(\Omega)$ finally reads
\begin{equation}
s(x) \, = \, \sum_{i \in \mathbb{Z}} \langle s ,  \widetilde \phi^{L}_{i} \rangle \, \phi^{L}_{i}(x) \, + \, \sum_{\ell = -\infty}^{L} \, \sum_{i \in
\mathbb{Z}} \langle s , \widetilde \psi^{\ell}_{i} \rangle \, \psi^{\ell}_{i}(x),
\label{eq:biorth_wl_series}
\end{equation}
which 
in the present case corresponds to the rescaled ($L^1$ normalized) Haar wavelet. In the case of finite resolution the finest scale $\ell= - \infty$ is typically replaced by $\ell=0$.
The special case of an orthogonal representation is recovered for $\widetilde \phi = \phi$ and $\widetilde \psi = \psi$, which means in our case renormalizing the basis functions in the $L^2$ norm.
Note that in general, for a given primary scaling function $\phi$ different dual scaling functions $\widetilde \phi $ can be constructed (and vise versa) and thus the choice is not unique.

For further details on this one-dimensional case using more general polynomial prediction operators, we refer, e.g., to the appendix of
\cite{deiterding2020multiresolution}.


Different normalizations of the basis functions are possible. In the context of cell average multiresolution typically a $L^1$ normalization is used. The $L^1$ norm of the dual scaling function $\widetilde \phi$ is equal to one, which is motivated by the use in the context of finite volume methods where cell averages are computed. For computing the divergence of the velocity of particle clouds using tesselations we are likewise using a finite volume approach and hence we adopted the $L^1$ normalization here.


\subsection{Case of graphs}

Starting with the projection and predition operators a similar connection between the discrete multiresolution analysis and biorthogonal wavelets on graphs can be established.
Using piecewise constant functions for each tesselation volume, a level dependent representation can be constructed, which is similar to Haar wavelets.
Wavelets are then introduced for representing the differences between two levels, either weighted or not. The corresponding filter coefficients can be likewise identified but they are position and level dependent.

We consider a signal field $s \in L^2(\Omega)$ in the continuous setting and the signal value 
at a discrete particle position is given by 
$\overline{s}_i^0 = \langle s, \widetilde{\phi}_{i}^0 \rangle$ ($i=0, \cdots, N^0-1$),
which represents the field value in the tessellation cell ${\Omega}^0_i$ with the volume $V_i^0$ assigned to the particle. 
The signal field $s({\bm x})$ with ${\bm x} \in \Omega$ can then be expressed as
\begin{equation}
    s^0({\bm x}) = \sum_{i=0}^{N^0-1} \overline{s}_i^0 \phi_{i}^0({\bm x}).
\end{equation}
The biorthogonal wavelet expansion is
\begin{equation}
    s^0({\bm x}) 
    = \sum_{i=0}^{N^L-1} \langle s, \widetilde{\phi}_{i}^L \rangle \phi_{i}^L({\bm x}) 
    + \sum_{\ell=1}^{L} \sum_{i=0}^{N^\ell-1} \langle s, \widetilde{\psi}_{i}^\ell \rangle \psi_{i}^\ell({\bm x})
\end{equation}
with
\begin{eqnarray}
  \overline{s}_i^\ell &=& \langle s, \widetilde{\phi}_{i}^\ell \rangle \\
  d_i^\ell &=& \langle s, \widetilde{\psi}_{i}^\ell \rangle 
\end{eqnarray}

In the multiresolution technique proposed in this paper, the primary scaling function 
$\phi_i^0({\bm x})$ and its dual 
$\widetilde{\phi}_i^0({\bm x})$ are defined by piecewise constant functions,
\begin{equation}
   \phi_{i}^0({\bm x}) 
   = \left\{ 
   \begin{array}{cc}
   1 & {\rm for} \; {\bm x} \in \Omega^0_i,  \\
   0 & {\rm otherwise},
  \end{array}
  \right.
   \label{eq:phi_0}
\end{equation}
and
\begin{equation}
   \widetilde{\phi}_{i}^0({\bm x}) 
   = \left\{ 
   \begin{array}{cc}
   1/V_i^0 & {\rm for} \; {\bm x} \in \Omega^0_i,  \\
   0       & {\rm otherwise}.
  \end{array}
  \right.
   \label{eq:phi_tilde_0}
\end{equation}

The scaling functions, 
$\phi_{i}^{\ell+1}({\bm x})$ and $\widetilde{\phi}_{i}^{\ell+1}({\bm x})$, 
for the level $\ell+1$ are 
given by those at the level $\ell$ as 
\begin{equation}
   \phi_{i}^{\ell+1}({\bm x}) 
   = \phi_{2i}^{\ell}({\bm x}) + \phi_{2i+1}^{\ell}({\bm x})
   \label{eq:phi_ell},
\end{equation}
and
\begin{equation}
   \widetilde{\phi}_{i}^{\ell+1}({\bm x}) 
   =  \frac{1}{V_{i}^{\ell+1}} \left\{
     V_{2i}^\ell \widetilde{\phi}_{2i}^{\ell}({\bm x}) 
   + V_{2i+1}^\ell \widetilde{\phi}_{2i+1}^{\ell}({\bm x}) 
   \right\}
   \label{eq:phi_tilde_ell},
\end{equation}
for $\ell = 0, \cdots, L-1$.

The wavelets, $\psi_{i}^{\ell+1}({\bm x})$ and $\widetilde{\psi}_{i}^{\ell+1}({\bm x})$, are then given by the scaling functions as
\begin{equation}
   \psi_{i}^{\ell+1}({\bm x}) 
   = V_{2i+1}^\ell \left\{ 
   \widetilde{\phi}_{2i+1}^{\ell}({\bm x}) - \widetilde{\phi}_{2i}^{\ell}({\bm x})
   \right\}
   \label{eq:psi_ell},
\end{equation}
and 
\begin{equation}
   \widetilde{\psi}_{i}^{\ell+1}({\bm x}) 
   = \frac{V^\ell_{2i}}{V^{\ell+1}_i} \left\{ \widetilde{\phi}_{2i+1}^{\ell}({\bm x}) - \widetilde{\phi}_{2i}^{\ell}({\bm x}) \right\}
   \label{eq:psi_tilde_ell},
\end{equation}
for $\ell = 0, \cdots, L-1$. 

The above wavelets and scaling functions satisfy the following relationships:
\begin{eqnarray}
   \int_{\Omega} \phi_{i}^\ell({\bm x}) d{\bm x} &=& V_i^\ell \\
   \int_{\Omega} \widetilde{\phi}_{i}^\ell({\bm x}) d{\bm x} &=& 1 \\
   \int_{\Omega} \psi_{i}^\ell({\bm x}) d{\bm x} &=& 0 \\
   \int_{\Omega} \widetilde{\psi}_{i}^\ell({\bm x}) d{\bm x} &=& 0 \\
   \langle \phi_i^\ell, \widetilde{\phi}_{i'}^{\ell} \rangle &=& \delta_{i i'} \\
   \langle \psi_i^\ell, \widetilde{\psi}_{i'}^{\ell'} \rangle &=& \delta_{i i'}\delta_{\ell \ell'} \\
   \langle \phi_i^\ell, \widetilde{\psi}_{i'}^{\ell'} \rangle &=& 0 \ {\rm for} \ \ell \ge \ell'
\end{eqnarray}


\section{Scaling factor for $L^2$ normalization}
\label{app2}

To obtain the scaling factor for $L^2$ normalization, we rewrite the wavelets and scaling functions using the scaling factors.
The primary scaling functions are described as
\begin{equation}
   \phi_{i}^0({\bm x}) 
   = \left\{ 
   \begin{array}{cc}
   1/\alpha_i^0 & {\rm for} \; {\bm x} \in \Omega^0_i,  \\
   0 & {\rm otherwise},
  \end{array}
  \right.
   \label{eq:phi_0}
\end{equation}
and the dual ones as
\begin{equation}
   \widetilde{\phi}_{i}^0({\bm x}) 
   = \left\{ 
   \begin{array}{cc}
   \alpha_i^0/V_i^0 & {\rm for} \; {\bm x} \in \Omega^0_i,  \\
   0       & {\rm otherwise}.
  \end{array}
  \right.
   \label{eq:phi_tilde_0}
\end{equation}

The scaling functions, 
$\phi_{i}^{\ell+1}({\bm x})$ and $\widetilde{\phi}_{i}^{\ell+1}({\bm x})$,
for the level $\ell+1$ are 
given by those at the level $\ell$ as 
\begin{equation}
   \phi_{i}^{\ell+1}({\bm x}) 
   = \frac{1}{\alpha_i^{\ell+1}} \left\{ \alpha_{2i}^{\ell} \phi_{2i}^{\ell}({\bm x}) + \alpha_{2i+1}^{\ell} \phi_{2i+1}^{\ell}({\bm x}) \right\}
   \label{eq:phi_ell},
\end{equation}
and
\begin{equation}
   \widetilde{\phi}_{i}^{\ell+1}({\bm x}) 
   =  \frac{\alpha_i^{\ell+1}}{V_{i}^{\ell+1}} \left\{
     \frac{V_{2i}^\ell}{\alpha_{2i}^{\ell}} \widetilde{\phi}_{2i}^{\ell}({\bm x}) 
   + \frac{V_{2i+1}^\ell}{\alpha_{2i+1}^{\ell}} \widetilde{\phi}_{2i+1}^{\ell}({\bm x}) 
   \right\}
   \label{eq:phi_tilde_ell},
\end{equation}
for $\ell = 0, \cdots, L-1$. Here $\alpha_i^{\ell+1}$ is the scaling factor for the scaling functions.

The wavelets, $\psi_{i}^{\ell+1}({\bm x})$ and $\widetilde{\psi}_{i}^{\ell+1}({\bm x})$, are then given by the difference of two 
scaling functions as
\begin{equation}
   \psi_{i}^{\ell+1}({\bm x}) 
   = \frac{V_{2i+1}^\ell}{\beta_i^{\ell+1}} \left\{ 
   \frac{1}{\alpha_{2i+1}^\ell} \widetilde{\phi}_{2i+1}^{\ell}({\bm x}) - \frac{1}{\alpha_{2i}^\ell} \widetilde{\phi}_{2i}^{\ell}({\bm x})
   \right\}
   \label{eq:psi_ell},
\end{equation}
and 
\begin{equation}
   \widetilde{\psi}_{i}^{\ell+1}({\bm x}) 
   = \frac{\beta_i^{\ell+1} V_{2i}^\ell}{V_{i}^{\ell+1}} \left\{ 
   \frac{1}{\alpha_{2i+1}^\ell} \widetilde{\phi}_{2i+1}^{\ell}({\bm x}) - \frac{1}{\alpha_{2i}^\ell}\widetilde{\phi}_{2i}^{\ell}({\bm x}) 
   \right\}
   \label{eq:psi_tilde_ell},
\end{equation}
for $\ell = 0, \cdots, L-1$. $\beta_i^{\ell+1}$ is the scaling factor for the wavelets.
By choosing $\alpha_i^\ell = \sqrt{V_i^\ell}$ and 
$\beta_i^{\ell+1} = \sqrt{V_i^{\ell+1} V_{2i+1}^\ell/V_{2i}^\ell}$, 
the scaling functions and wavelets for the $L^2$ normalization are obtained as
\begin{equation}
   \phi_{i}^0({\bm x}) = \widetilde{\phi}_{i}^0({\bm x}) 
   = \left\{ 
   \begin{array}{cc}
   1/\sqrt{V_i^0} & {\rm for} \; {\bm x} \in \Omega^0_i,  \\
   0 & {\rm otherwise},
  \end{array}
  \right.
   \label{eq:phi_0}
\end{equation}
\begin{equation}
   \phi_{i}^{\ell+1}({\bm x}) =\widetilde{\phi}_{i}^{\ell+1}({\bm x}) 
   = \frac{1}{\sqrt{V_i^{\ell+1}}} \left\{ \sqrt{V_{2i}^{\ell}} \phi_{2i}^{\ell}({\bm x}) + \sqrt{V_{2i+1}^{\ell}} \phi_{2i+1}^{\ell}({\bm x}) \right\}
   \label{eq:phi_ell},
\end{equation}
\begin{equation}
   \psi_{i}^{\ell+1}({\bm x}) = \widetilde{\psi}_{i}^{\ell+1}({\bm x}) 
   = \sqrt{\frac{V_{2i}^\ell}{V_{i}^{\ell+1}}} \widetilde{\phi}_{2i+1}^{\ell}({\bm x}) 
   - \sqrt{\frac{V_{2i+1}^\ell}{V_{i}^{\ell+1}}} \widetilde{\phi}_{2i}^{\ell}({\bm x})
   \label{eq:psi_ell},
\end{equation}
for $\ell = 0, \cdots, L-1$. These scaling functions and wavelets satisfy 
$\langle \phi_i^\ell, \phi_{i}^{\ell} \rangle = \langle \widetilde{\phi_i^\ell}, \widetilde{\phi}_{i}^{\ell} \rangle = \langle \psi_i^\ell, \psi_{i}^{\ell} \rangle = \langle \widetilde{\psi}_i^\ell, \widetilde{\psi}_{i}^{\ell} \rangle = 1$, i.e., the $L^2$ norm is equal to one.
Since the wavelet coefficients are given by $d_i^{\ell+1} = \langle s, \widetilde{\psi}_i^{\ell+1} \rangle$, the scaling factor of the wavelet coefficient for $L^2$ normalization is $\sigma_i^{\ell}=\beta_i^{\ell}$.



\bibliographystyle{elsarticle-num} 
\bibliography{reference}



%
%
%
\end{document}